\documentclass[a4paper,fleqn,usenatbib]{mnras}

\usepackage{newtxtext,newtxmath}

\usepackage[T1]{fontenc}
\usepackage{ae,aecompl}
\usepackage{xspace}
\usepackage{times}
\usepackage{amsmath} 
\usepackage{amssymb} 
\usepackage{graphicx} 
\usepackage{indentfirst} 
\usepackage{pdflscape} 
\usepackage{placeins} 
\usepackage{nicefrac} 
\usepackage{afterpage} 
\usepackage{rotating} 
\usepackage{booktabs} 
\usepackage{rotating} 
\usepackage{color} 
\usepackage{caption} 
 \usepackage{xprintlen}
 \usepackage[T1]{fontenc}
\usepackage{ae,aecompl}

\usepackage{graphicx}	
\usepackage{amsmath}	
\usepackage{amssymb}	

 \newcommand{\nab}{\mbox{\boldmath$\nabla$}}
 \newcommand{\vel}{\mbox{\boldmath$v$}}

\hypersetup{draft} 

\title[The circumgalactic mist]{On the model of the circumgalactic mist: the implications of cloud sizes in galactic winds and halos}

\author[Liang \& Remming]{
Cameron J. Liang$^{1,2}$\thanks{E-mail:jwliang@oddjob.uchicago.edu}\thanks{NASA Earth \& Space Science Fellow}
and Ian Remming$^{1}$
\\
$^{1}$Department of Astronomy \& Astrophysics, The University of Chicago, Chicago, IL 60637, USA\\
$^{2}$Kavli Institute for Cosmological Physics, The University of Chicago, Chicago, IL 60637, USA\\
}

\date{Accepted XXX. Received YYY; in original form ZZZ}

\pubyear{2018}

\begin{document}
\label{firstpage}
\pagerange{\pageref{firstpage}--\pageref{lastpage}}
\maketitle

\begin{abstract}
Ubiquitous detections of cold/warm gas around galaxies indicate that the circumgalactic medium (CGM) is multiphase and dynamic. Recent state-of-the-art cosmological galaxy simulations have generally underproduced the column density of cold halo gas. We argue that this may be due to a mismatch of spatial resolution in the circumgalactic space and the relevant physical scales at which the cold gas operates.  Using semi-analytic calculations and a set of magnetohydrodynamic (MHD) simulations, we present a multiphase model of the gaseous halos around galaxies, the circumgalactic mist (CGm). The CGm model is based on the idea that the observed cold halo gas may be a composite of cold, dense and small cloudlets embedded in a hot diffuse halo, resembling terrestrial clouds and mist. We show that the resulting cold gas from thermal instabilities conforms to a characteristic column density of $N_{\rm H}\approx 10^{17}\rm{cm^{-2}}$ as predicted by the $c_s t_{\rm cool}$ ansatz. The model implies a large number of cold clumps in the inner galactic halo with a small volume filling factor but large covering fraction. The model also naturally gives rise to spatial extents and differential covering fractions of cold, warm and hot gas. To self-consistently model the co-evolution of the CGM and star formation within galaxies, future simulations must address the mismatch of the spatial resolution and characteristic scale of cold gas. 
\end{abstract}




\section{Introduction}


The conceptualization of gaseous halos around galaxies arose from a hypothesis by \cite{Spitzer1956}: the observed cold gas at large distances away from the Milky-Way galactic plane \citep{MunchZirin1961} must be pressure supported by a hot corona.  This idea was later incorporated into the  Lambda Cold Dark Matter ($\Lambda$CDM) paradigm where intergalactic gas accretes onto dark matter potential wells and is shocked heated to the virial temperature of the halo \citep[e.g.,][]{WhiteRees1978}. Radiative cooling is thought to play an important role subsequently \citep{Binney1977, Silk1977, ReesOstriker1977} and leads to multiphase and dynamic gaseous halos \citep{MoMiraldaEscude1996, MallerBullock2004}.


Observational evidence of multiphase galactic winds and halo gas has accumulated over the years in support of this picture.  In the early days, the Mg\,II doublet $\lambda\lambda$2976, 2803 has primarily been used for probing the connection of heavy elements with host galaxies at moderate redshifts \citep[e.g.,][]{BergeronBoisse1991, Churchill1996, Steidel2002, Chen2010}.  This is owing to the strong oscillator strength and relatively long wavelength of the Mg\,II doublet where it can be observed in the optical regime at intermediate redshifts.  At the same time, galaxies with similar redshifts can be observed relatively easily using ground-based telescopes, such as the Sloan Digital Sky Survey \citep[SDSS;][]{York2000}.


The launch of the \textit{Hubble Space Telescope} (HST) and the Cosmic Origin Spectrograph \citep[COS;][]{Green2012} brought a renaissance to the galaxy-gaseous halo connection and characterizations.  The ultraviolet (UV) wavelength coverage and the high sensitivity of COS allow one to measure absorption lines of a broad range of ions, probing a wide range of physical conditions of the multiphase CGM.  These include early studies probing neutral hydrogen using Ly$\alpha$ $\lambda$1215 \citep{Lanzetta1995, Chen1998, Chen2001}, heavy elements with C\,IV $\lambda\lambda$ 1548,1550 \citep{Chen2001b},  and highly ionized O\,VI \citep{ChenMulchaey2009, Prochaska2011}.  

In recent years, there has been an accelerated progress thanks to a combination of large HST and ground-based observing programs that probe a broad range of ions at once (e.g., H\,I, Si\,II, Si\,III, Si\,IV, C\,II, C\,IV, O\,VI). For example, the COS-Halo program \citep{Tumlinson2011, Werk2013} probes a set of star-forming and quiescent galaxies within 160 kpc.  The authors show the CGM harbors a significant fraction of the baryons  \citep{Werk2014}.  The COS-GASS program explores the connection between the interstellar medium (ISM) and the CGM by combining with HI 21 cm line observations \citep{Borthakur2015}.  While these studies focus on the inner regime of the CGM of some specific types of galaxies,  \cite{LiangChen2014}, in contrast, used 195 galaxy-QSO pairs to probe CGM with a broad range of stellar mass of the galaxies ($10^7 - 10^{12} M_{\odot}$) and out to large distances (500 kpc). With a stringent limit, they show that detection rates of low ions (e.g., Si\,II) drop quickly beyond a small fraction of the virial radius ($0.4-0.5 R_{\rm vir}$), while intermediate ions (e.g., Si\,IV, C\,IV)  are detected up to $\sim 0.7 R_{\rm vir}$.  \cite{Johnson2015} extended this study to highly ionized gas where they found continuous detection of O\,VI up to $\sim R_{\rm vir}$. 

One of the natural explanations for this enrichment of the CGM is via galactic winds \citep[e.g.,][]{Muratov2015}. Galactic outflows are commonly observed in both the local and high-redshift Universe \citep[e.g.,][]{Shapley2003, Veilleux2005}. In addition to the hot phase observed through X-ray emitting gas \citep[e.g.,][]{StricklandHeckman2009}, the outflows can be observed via self-absorption of Na\,I$\lambda$ 5890,5896 and Mg\,II doublets \citep[e.g.,][]{Heckman2000, Weiner2009} and warm ionized gas \citep[e.g.,][]{Lehnert1999,Heckman2017}. In some cases, even molecular gas can also be detected \citep[e.g.,][]{Veilleux2009,Beirao2015,Leroy2015}.


These analyses have provided a comprehensive characterization of multiphase gas in a broad range of galaxies.  The task from the theoretical side has been to understand the implications of these observations and incorporate the governing physical processes into our modeling of galaxy formation.  The current generation of galaxy formation models appears to capture many of the relevant physical processes. State-of-the-art simulations can now produce galaxies with realistic stellar properties \citep{Feldmann2011, Sales2012,Aumer2013,Stinson2013, Hopkins2014, AgertzKravtsov2015, Schaye2015}. This success in the theoretical community can be mainly attributed to our improved understanding of star formation and feedback processes, and the corresponding improved subgrid recipes in our simulation codes.  A general conclusion is that galactic winds are believed to play crucial roles in shaping the galaxy as a whole. It has been shown that star formation feedback can drive large-scale galactic winds to regulate star formation. In doing so, the hot winds can provide the necessary heating to balance the global cooling of gaseous halos. 

Despite the success of recent simulations in producing realistic stellar properties of galaxies,  matching the CGM properties proves to be difficult.   Simulations can explain some CGM properties, such as the bimodality of O\,VI column densities and specific star formation rate (sSFR) of the host galaxies \citep{Tumlinson2011, Oppenheimer2016}.  However, most simulations have been unable to produce enough cold/warm gas in the CGM, despite a diversity of feedback processes implemented in them \citep[e.g.,][]{Stinson2012, Ford2013, Hummels2013, Shen2013, suresh2015}. On the other hand,  a variation of large feedback energy in \cite{Liang2016} leads to a more realistic CGM by launching cold gas into the halo; however it does so by destroying their central galaxy. It is also thought that cosmic rays feedback may help launch the cold gas into large distances \citep[e.g.,][]{Socrates2008, Booth2013, Salem2014}, although it has been shown that these initial attempts were not yet enough \citep[e.g.,][]{Liang2016, Salem2016,ButskyQuinn2018}.  

We are motivated by many factors to consider new solutions to the problem of underproducing the cold/warm phase of the CGM in simulations. These factors include (1) the ubiquitous detections of cold gas in halos of nearly all types of galaxies, (2) small sizes of cold gas, which are both observationally and theoretically motivated, and (3) the common underproduction despite a broad range of simulation codes, star formation and feedback processes and implementations.  In fact, it is becoming clear that galaxy simulations with cosmological volume lack the appropriate spatial resolution to resolve the sizes of cold gas. This is because these simulations traditionally focus their computational power on refining the densest region of galaxies. As a result, the galactic halo of low-density gas is resolved only down to the order of $\sim$ kpc.  By contrast, it has been shown observationally that the cold gas is in the order of $\sim$ pc \citep[e.g.,][]{Rauch1999, ProchaskaHennawi2009, Crighton2015, LanFukugita2017}. Similarly, theoretical models and small-scale simulations focused on the galactic winds and thermal instabilities have long predicted pc-size cold cloudlets under various conditions \citep[e.g.,][]{BurkertLin2000, SanchezSalcedo2002, AuditHennebelle2005, Cooper2009, McCourt2018}.  Models that consider condensation of cold gas from hot halo have also been quite successful in explaining observations of intracluster medium \citep[e.g.,][]{McCourt2012, Gaspari2012, Li2015, Voit2015b, Gaspari2017, Fogarty2017}

In summary, the challenges and tantalizing hints from this line of observational and theoretical studies point towards the need to consider small-scale structure of cold gas in the circumgalactic context. Thus, we explore a new multiphase model, the circumgalactic mist or CGm, and its implications for the multiphase nature of galactic winds and gaseous halos.  In a series of papers, we aim to investigate the CGm model on explaining a multitude of observations, including the cloud sizes, the gas content at various phases, their kinematics, and more.

In this paper, we focus on the production of cold gas due to thermal instability in a set of small-scale simulations with an array of physical processes to study the resulting cloudlet sizes. We are interested in the spatial distribution the CGM in various phases, in particular, the presence of observed small cold clouds in the inner CGM, and the lack thereof at larger radii. We leave the survival of cold cloudlets and the kinematics of the circumgalactic mist in a future study. 

The structure of the paper is organized as follows. In section \ref{sec:rclmodel}, we discuss the process of thermal instability due to runaway radiative cooling with a focus on the resulting sizes of cold clumps. We provide a simple semi-analytic model to describe the evolution of isobaric cooling and cloudlet sizes. In section \ref{sec:rclsim}, we use a set of MHD simulations to study an episode of the birth and death of cold gas and their sizes in the presence of a hot wind. We systematically introduce physical processes, such as radiative cooling/heating, thermal conduction and magnetic fields to study their effects on the cloudlet sizes or the lack thereof.  In section \ref{sec:rclcgm}, we discuss the implication of the characteristic cloudlet sizes predicted by the model, showing that the inner CGM may be composed of an upward of millions of cold clumps with sizes in the order of parsecs. Consequently, the model also predicts the spatial extent of ions probing the cold, warm and hot phase of the CGM.  Throughout this paper, we have used the term cloudlet as a shorthand for characteristic parsec-sized regions of cold gas with increased density.

\section{Semi-Analytic Consideration}
\label{sec:rclmodel}

Much of the previous analyses of thermal instability has focused on the process of runaway cooling that leads to a multiphase medium at a final equilibrium state \citep[e.g.,][]{Parker1953, Field1965, HennebellePerault1999, BurkertLin2000}.  Famous two-phase and three-phase models of the interstellar medium (ISM) were born out of the studies of thermal instabilities \citep{FieldGoldsmithHabing1969, McKeeOstriker1977}.  Cloud formation and compressible turbulence due to isochoric and isobaric cooling have been an area of active research for many years as well \citep[e.g.,][]{ KritsukNorman2002, SanchezSalcedo2002, AuditHennebelle2005}. 

We begin our study with theoretical expectations based on semi-analytic calculations. In this section, we consider the nonlinear evolution of thermal instability via fragmentation and monolithic condensation. However, we are most interested in the final characteristic size of cloudlets from which thermal instabilities bring.

\subsection{Fragmentation in thermal instability}

\begin{figure}
\begin{center}
\hspace{30mm}
\includegraphics[scale=0.7]{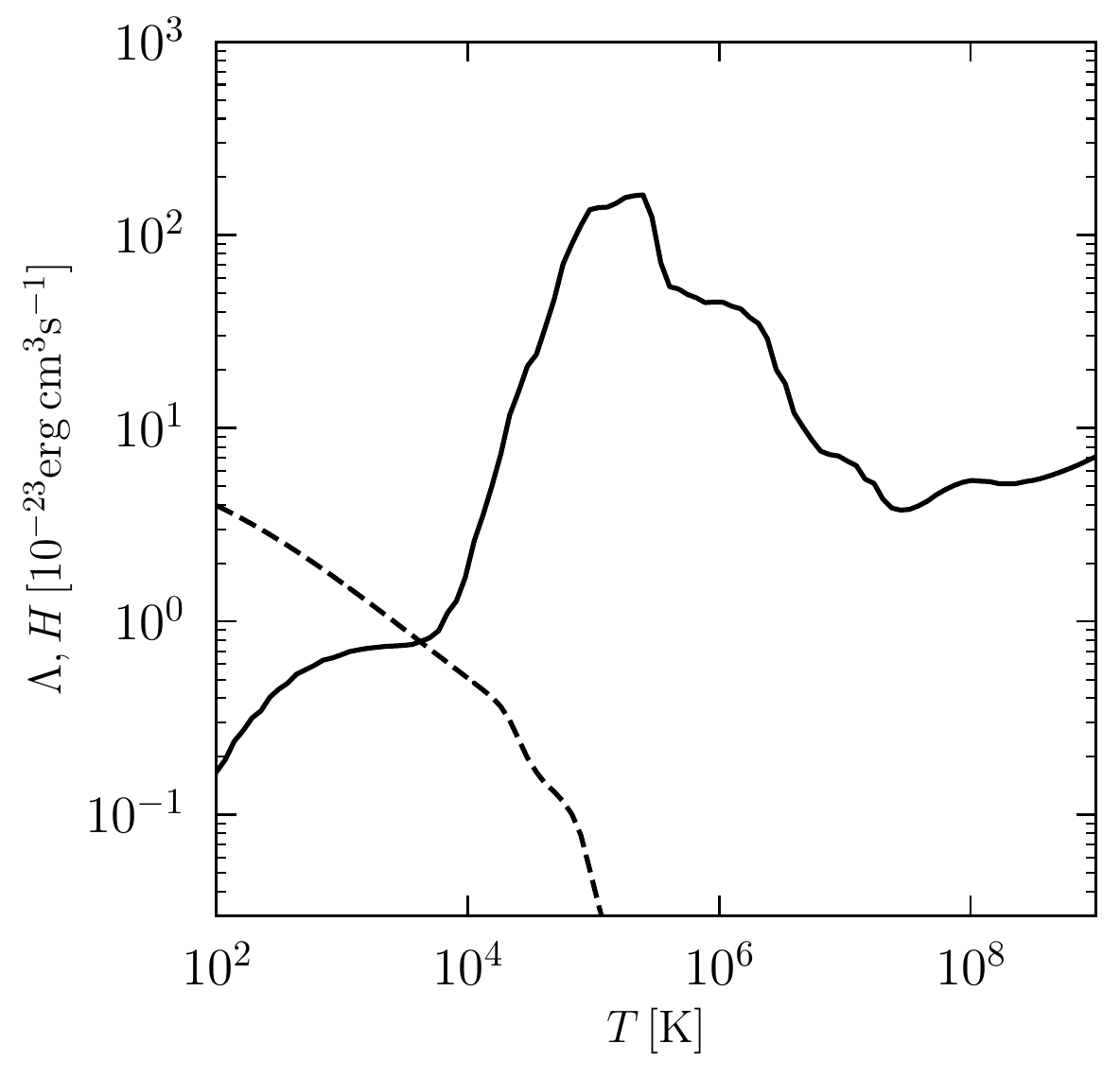}
\caption{Cooling (solid) and heating (dashed) curves used throughout this work at $Z/Z_{\odot} = 0.3$ \citep{GnedinHollon2012}. \label{fig:coolfunc}}
\end{center}
\end{figure}

\begin{figure}
\begin{center}
\hspace{30mm}
\includegraphics[width=\columnwidth]{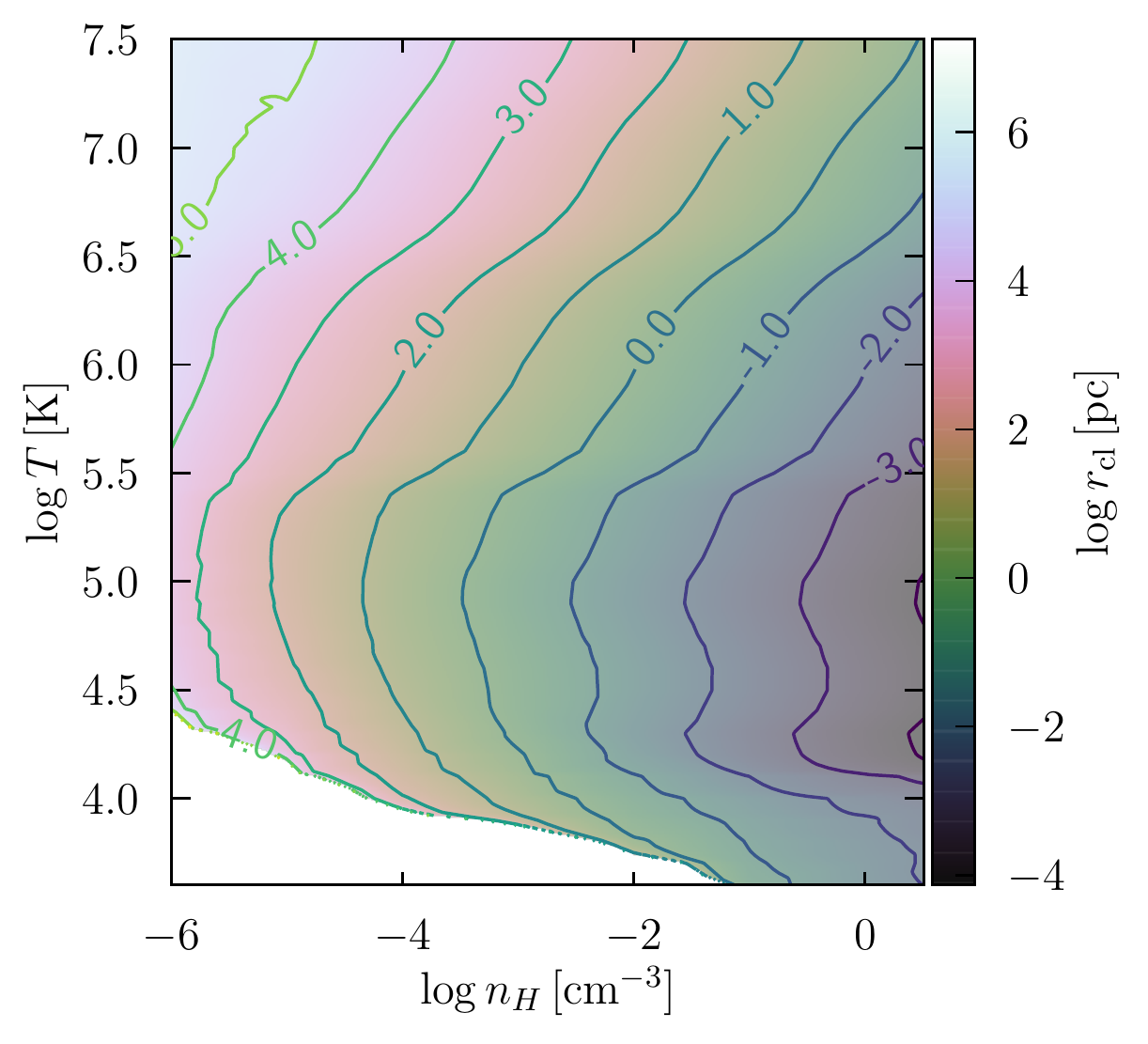}
\caption{The size of cloudlets under the ansatz, $r_{\rm cl} = c_s (T) t_{\rm cool} (n, T, Z)$, as a function of density and temperature. The contours are constant cloudlet sizes in parsecs at $Z = 0.3 Z_{\odot}$. The contours of different sizes converge to the lower boundary, which is set by the balance of heating and cooling. In other words,  the constraint of the cloudlet size disappears as $t_{\rm cool} \rightarrow \infty$. At this point, the problem becomes scale-free. Note that at the typical condition of observed/warm gas ($T<10^{4.5} $K and $n_{\rm H} \sim 10^{-3} \rm{cm^{-3}} $), the expected cloudlet size is of the order of 1 pc or less.  \label{fig:rmap}}
\end{center}
\end{figure}

It is well known that a region of perturbed density of an initial scale $r_{\rm{cl},i}$ will undergo condensation if the cooling time is large compared to the corresponding sound crossing time, $t_{\rm cool}/t_{\rm cross} >1$ \citep[e.g.,][]{Field1965, BurkertLin2000}.   In contrast, regions of density perturbations with $t_{\rm cool}/t_{\rm cross} <1$ cool more quickly than thermal motion can equilibrate pressure; large pressure gradients can develop. In turn, these pressure forces lead to a more dynamic evolution of the condensation \citep[e.g.,][]{Field1965, Meerson1996, SanchezSalcedo2002}.

Most recently, \cite{McCourt2018} argued that a faster way to reach the equilibrium state is via fragmentation, similar to the process of the Jeans' instability in gravitational collapse. This thermal fragmentation process ensues when the communication across a cloudlet by thermal motion is slow compared to the cooling time $t_{\rm cool}$. In other words, this shattering process stops when thermal pressure is able to smooth out perturbations faster than their growths: 

\begin{equation} \frac{r_{\rm cl}}{c_s} \equiv  t_{\rm cross} \leq t_{\rm cool}  \end{equation}
where $r_{\rm cl}$ is the size of the perturbed density region (a cloudlet) and $c_s$ is the sound speed. 

Therefore, one can write down the expected cloudlet size at the end of the fragmentation process based on this ansatz.  Given a metallicity $Z_{\rm cl}$, sizes of cloudlets live on a density-temperature phase space (Figure \ref{fig:rmap}) that satisfy the following condition: 

\begin{equation} r_{\rm cl} \leq c_s (T_{\rm cl}) t_{\rm cool} (n_{\rm cl}, T_{\rm cl}, Z_a) \label{eqn:cstcool} \end{equation} 
The equality predicts the allowed cloudlet sizes at an evolutionary boundary between fragmentation and monolithic collapse as a function of the temperature, density, and metallicity. We can see in Figure \ref{fig:rmap} that the cloudlet sizes can be very small. At a typical physical condition of the observed cold/warm gas with $T<10^{4.5} $K and $n_{\rm H} \sim 10^{-3} \rm{cm^{-3}} $, the expected size is in the order of 1 pc or less. However, it is worth pointing out that the growth rate is a function of spatial scale. In fact, the perturbation can only grow as fast as pressure gradient pushes them around on a time scale of $t_{\rm cross}$. 

The contours of constant cloudlet sizes in Figure \ref{fig:rmap} converge to a lower boundary set by the balance of heating and cooling. At this locus, the cooling time is so long that it no longer limits the size of the cloudlet (i.e., $t_{\rm cool} \rightarrow \infty$). 

Models of cloud formation that describe the size of cold gas have mainly focused on the competition of cooling time and sound crossing time at the hot phase \citep[e.g.,][]{HennebellePerault1999, BurkertLin2000}. As noted by \cite{McCourt2018}, the difference here is that the final cloudlet size depends on the cooling time and the sound speed evaluated near the temperature at which $c_s t_{\rm cool}$ is minimized: 

\begin{equation} r_{\rm cl} \approx \rm{min} (c_s t_{\rm cool})  \label{eqn:rclmin}\end{equation}

Figure \ref{fig:r_tp} shows the length scale of $c_s t_{\rm cool}$ as a function of temperature. The corresponding temperature at which Eq. \ref{eqn:rclmin} is minimized happens at $T_{\rm min} \approx 10^{4.3}$ K.  It is worth pointing out that the parcel of dense gas will continue to cool as long as its cooling time is not formally infinite. However, the increase of cooling time is dramatic with $T < T_{\rm min}$ in the regime we are considering. Therefore, we assume the cooling and density growth are negligible after this point; the final temperature of the cloudlet is close to this minimum temperature, $T_{\rm cl,f} \approx T_{\rm min}$, as we will discuss in what follows.

\subsection{Isobaric condensation mode in thermal instability}

\begin{figure}
\begin{center}
\includegraphics[width=\columnwidth]{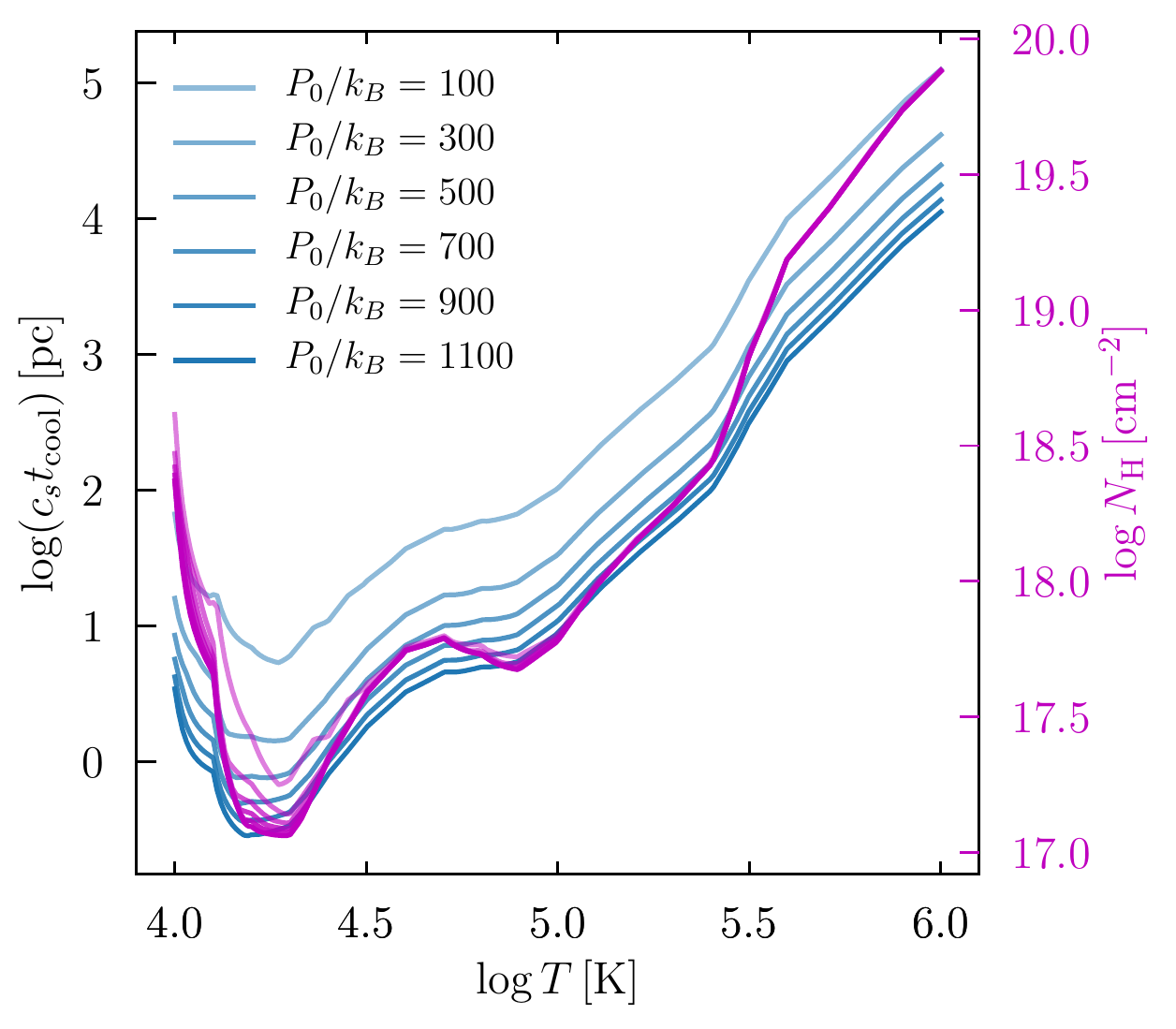}
\caption{The cloudlet sizes as function of temperature for different ambient pressures. The gradient of blue lines shows higher background pressure is able to compress cloudlets more, leading to smaller sizes. The corresponding column densities of cloudlets are represented by the series of magenta lines of different opacities. As shown, the column densities are insensitive to the change of background pressure.   \label{fig:r_tp}}
\end{center}
\end{figure}

\begin{figure*}
\begin{center}
\hspace{30mm}

\includegraphics[width=17 cm,height=15 cm]{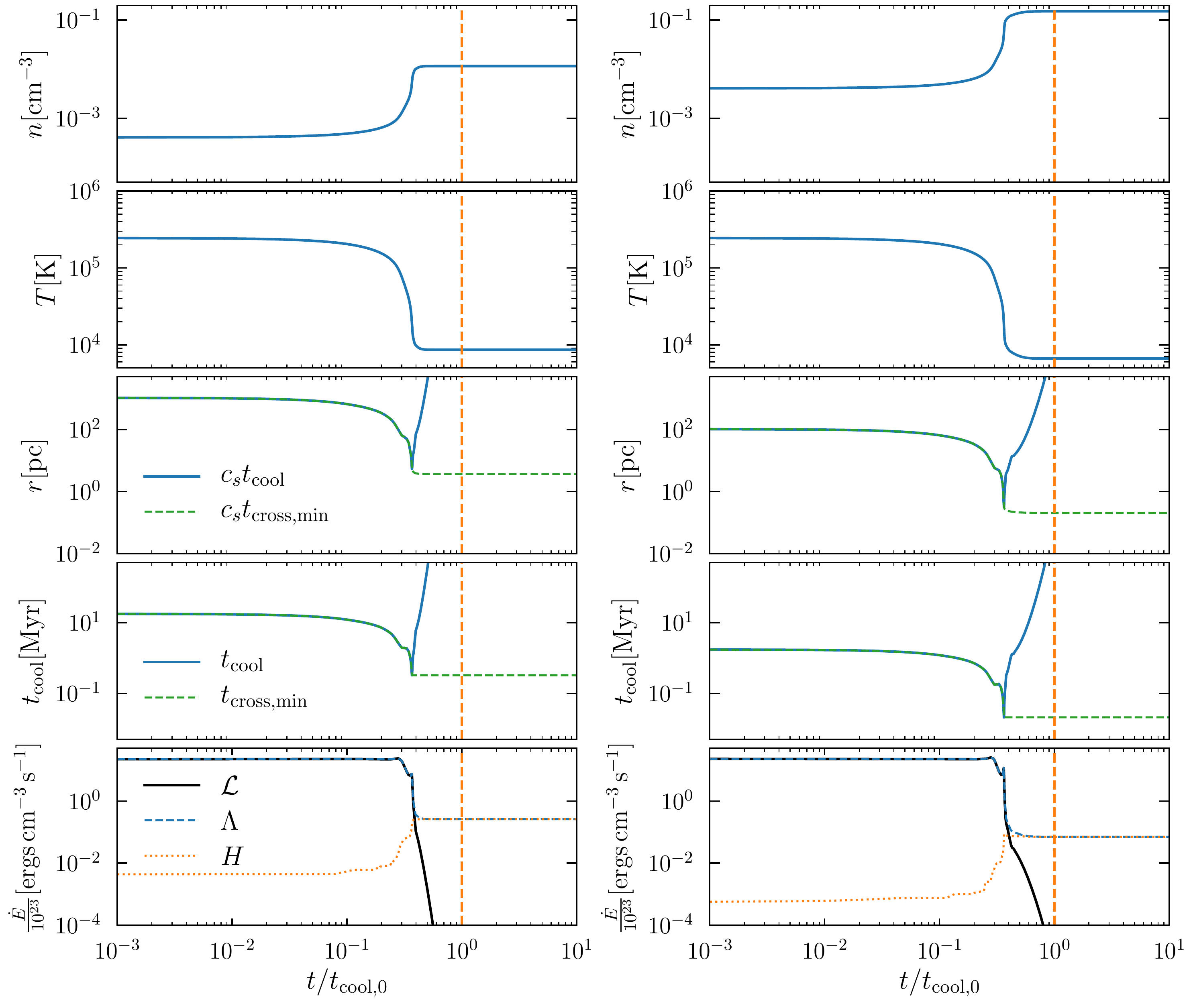}
\caption{The evolution of the cloudlet properties given the background pressure with $P/k_{\rm B} = 100$ (left panel) and $P/k_{\rm B} = 1000$ (right panel). As shown, cloudlet quickly cools and condenses within a cooling time of the initial state $t_{\rm cool,0}$. Upon reaching the minimum of the cooling time, the radius of the cloudlet follows $r_{\rm cl} = c_s t_{\rm cross,min}$ as $t_{\rm cool}$ departs and grows as the net cooling rate approaches zero.  The final size of the cloudlet depends on the external pressure as shown, where $r_{\rm cl} \approx 0.2$ pc for $P/k_{\rm B} = 1000$, and $r_{\rm cl} \approx 3$ pc for $P/k_{\rm B} = 100$. \label{fig:cloudevolve}}
\end{center}
\end{figure*}

 In this subsection, we study the evolution of thermal instability by considering an initial uniform perturbed density region with a size $r_{\rm{cl},i}$, such that with $t_{\rm cool} = t_{\rm cross}$. In this regime,  the perturbed density will undergo isobaric condensation. 

While the Jeans' instability is driven by the local gravity of the collapsing objects, the size of gaseous cloudlets due to runaway cooling depends on the competition between its internal pressure and the ambient pressure.  As a region of overdensity cools, internal pressure support is lost. The ambient pressure (if higher) squeezes the overdensity further to maintain pressure equilibrium. This isobaric compression continues until heating and cooling are balanced, a pressure equilibrium state is reached at the end (e.g., see Figure \ref{fig:pequil}). 
From this, we can simply rewrite the final density of cloudlet as $n_{\rm cl,f} = P_{h} / (k_B T_{\rm cl,f})$, where $P_{h}$ is the pressure of the ambient hot gas. 

The final temperature $T_{\rm cl,f}$ and the minimum of the cooling time depend on the cooling function, via the metallicity of the gas in which the cloudlet condenses out of (i.e.,  $Z_{\rm cl} = Z_{h}$). Therefore, it is useful to express the cloudlet properties (density, temperature, and size) entirely as functions of their environment, $P_h$ and $Z_h$.

As discussed above, one can consider a cloudlet that satisfies the condition given by Eq. (\ref{eqn:cstcool}) with uniform perturbed density of size $r_{{\rm cl}, i}$ in an idealized setting. This idealized parcel of gas would not continue to shatter into smaller fragments in the absence of perturbations of smaller scales. Nevertheless, the cloudlet will still undergo runaway condensation due to radiative cooling ($\dot{e} = n^2 \mathcal{L}$) as long as there exists net cooling $\mathcal{L} = \Lambda - H > 0$. Using the fact that the final size of a cloudlet is tied to its thermal properties, we can solve for it by considering its energy evolution.  In this idealized setting, we assume thermal conduction is suppressed owing to the ubiquitous detections of cold gas in hot halos. Thus, the accounting of the thermal energy of the cloudlet is: 

\begin{equation} \frac{de_{\rm cl}}{dt} = - n_{\rm cl}^2 (\Lambda - H) + \frac{P_{\rm cl}}{n_{\rm cl}} \frac{dn_{\rm cl}}{dt}  \end{equation}

The first term on the right-hand side is the loss of internal energy by radiative cooling; the second term is the work done by the background through pressure forces. If we consider isobaric compression over a time interval $dt$, the internal energy of the cloudlet does not change due to the supply from the environment ($\frac{de_{\rm cl}}{dt} = 0$). This condition would be sustained if there is an adequate amount of heating (e.g., from galactic feedback processes) to balance the energy loss of the hot bath. 

The evolution of the cloudlet density thus follows this simple equation: 

\begin{equation} \frac{dn_{\rm cl}}{dt} = \frac{n_{\rm cl}^3}{P_{\rm cl}} (\Lambda - H)  \label{eqn:clt} \end{equation}

Formally, the cloudlet reaches its final state when it can no longer grow in density ($dn/dt = 0$). Equivalently, the equilibrium state is when the radiative loss approaches the photo-heating (i.e., $\mathcal{L} = \Lambda - H \rightarrow 0$).  At mentioned above, the subsequent growth in density is negligible after the cloudlet reaches a critical temperature since the cooling time grows steeply for $T < T_{\rm cl,f}$. Therefore, we simply can approximate the final size using Eq. (\ref{eqn:cstcool}) at $T_{\rm cl,f}$. In Figure \ref{fig:r_tp}, we show $c_s t_{\rm cool}$ as a function of temperature and ambient pressure. For the chosen metallicity at $Z = 0.3 Z_{\odot}$, the minimum size near the end of this evolutionary stage is at $T_{\rm cl,f} \approx 10^{4.3}$ K.  

Note the uptick of the size, in Figure \ref{fig:r_tp},  using $c_s t_{\rm cool}$ for $T < T_{\rm cl,f}$ no longer applies as $t_{\rm cool}$ departs from $t_{\rm cross}$. Obviously, there are no reasons for a cloudlet to expand so that its crossing time can continue matching its cooling time. In reality, the cloudlet has, by definition, a size of $r_{\rm cl} = c_s (T) t_{\rm cross}(T)$ as it cools. In a sense, we are exploiting the fact that the background hot gas attempts to reach pressure equilibrium in the shortest cooling time of a cloudlet by using Eq. (\ref{eqn:rclmin}), which allows one to  independently approximate $t_{\rm cross, min}$ with $t_{\rm cool,min}$.

In Figure \ref{fig:cloudevolve}, we show the evolution of thermodynamic properties of cloudlets by directly solving Eq. (\ref{eqn:clt}) given a background pressure and a fixed metallicity at $Z = 0.3Z_{\odot}$.  As shown, the cooling time (and thus $c_s t_{\rm cool}$) hovers at nearly the same value up to tens of percent of the initial cooling time $t_{\rm cool,0}$ when the density of the gas is small. As soon as the density grows large enough,  the bulk of cooling happens very quickly as cooling time drops precipitously. However, this stage of cooling does not last very long because the heating rate, which depends only on temperature, quickly approaches cooling rate at low temperature.

As shown in Figure \ref{fig:cloudevolve}, it is also clear that higher background gas pressure will lead to denser, colder and smaller cloudlets.  The final size is independent of the initial temperature and density of the cloudlet. It is also worth noting that the cloudlet properties reach the same equilibrium very quickly at about a cooling time $t_{\rm cool,0}$ evaluated at the initial state, even if the initial conditions are changed.

This model due to thermal fragmentation and/or isobaric cooling provides an interesting prediction: the column density of the cloudlet has a characteristic value at $N_{\rm H} \approx 10^{17} \rm{cm^{-2}}$ \citep[see also][]{VoitDonahue1990, McCourt2018}.  The minimum cloudlet size is $r_{\rm cl}  = c_s t_{\rm cool} \propto f(T)/n_{\rm cl}$, where $f(T)$ is the combination of the temperature dependence from the sound speed and cooling function.  The column density is $N_{\rm H} \approx n_{\rm cl} r_{\rm cl} \propto f(T)$. Therefore, as long as the temperature of the final state does not change very much with metallicity, the resulting column density will not differ dramatically. Using a realistic cooling curve at $Z = 0.3 Z_{\odot}$, we show that lines correspond to different pressures collapse onto nearly the same region, with minima between $10^{17} - 10^{17.5} \rm{cm^{-2}}$ (Figure \ref{fig:r_tp}). Note that the column density in Figure \ref{fig:r_tp} is calculated with the cloudlet diameter instead, $N_{\rm H} = 2 n_{\rm cl} r_{\rm cl}$, to compare cross section of cold gas in simulations in the section 3.

\section{Simulations}
\label{sec:rclsim}

\begin{figure}
\begin{center}
\hspace{30mm}
\includegraphics[scale=0.75]{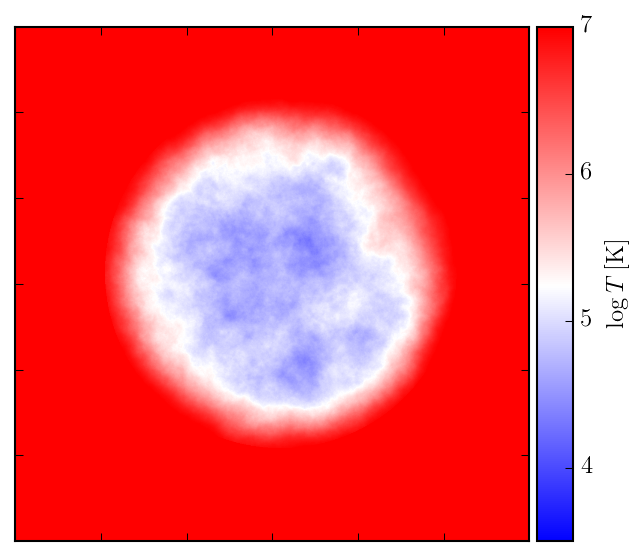}
\caption{Initial condition of the density region with isobaric density perturbation with radius $\sim 150$pc. The region has a median temperature of $\sim10^{5}$K and is tapered by a gaussian on the edges that smoothly transitions into the background temperature at $T_h = 10^7$K and density $n_{\rm h} = 10^{-4} \,\rm{cm^{-3}}$. \label{fig:IC}}
\end{center}
\end{figure}

In this section, we use a suite 10 two-dimensional MHD simulations to study an episode of birth and death of cold gas in the presence of hot winds in the circumgalactic context. We focus on the spatial scale of cold gas from a combination of thermal instabilities and the shock-cloud interactions. We also vary input physical processes systematically, such as cooling, (an)isotropic conduction, and magnetic fields, to study how they affect the resulting fragments of cold gas. We check our theoretical expectation of cloudlets by a comparison with their column densities across simulations. 
\subsection{Physics and initial conditions in simulations}

We use the High-Speed Combustion and Detonation (HSCD) code \citep{Khokhlov1998} for our simulations. The implementations of radiative cooling and adaptive mesh refinement are identical to those in the Adaptive Refinement Tree code \citep[ART;][]{Kravtsov1997}, which HSCD shares a common ancestor. We solve a set of compressible ideal MHD equations using a third order finite difference, weighted essentially non-oscillatory (WENO) scheme with magnetic field divergence cleaning \citep{Dedner2002, Mignone2010}. We also implement isotropic and anisotropic thermal conduction following the first-order super-time-stepping scheme as described in \cite{Meyer2014}. 

The effects of thermal conduction, radiative cooling and heating are described in the energy equation: 

 \begin{equation}
\frac{\partial e}{\partial t}+\nab \cdot \left(e + P + \frac{|\bmath{B}|^2}{8 \pi}\right)\vel - (\vel \cdot \bmath{B})\frac{\bmath{B}}{4 \pi}=  n^2 ( \Gamma - \Lambda ) +  \nab \cdot \bmath{F}_c \label{eqn:Eeqn}
\end{equation}

$\Gamma$ and $\Lambda$ are the heating and cooling functions. We adopt a nominal value of $Z = 0.3 Z_{\odot}$ for the cooling function for all simulations. The heating and cooling functions are shown in Figure \ref{fig:coolfunc} \citep{GnedinHollon2012}.  $\nab \cdot \bmath{F}_c$ is the net heat flux contribution from thermal conduction that smoothly transitions from the classical heat flux to the saturated regime: 

\begin{equation}
  \bmath{F}_c = \frac{F_{\text{sat}}}{F_{\text{sat}} + | \bmath{F}_{\text{class}} | } \bmath{F}_{\text{class}}
\end{equation}
where $F_{\rm sat}$ is the saturated heat flux when the mean free path of electrons is large compare to the local temperature scale length such that the transfer of heat is less efficient. Following \cite{CowieMcKee1977}, we compute $F_{\rm sat}$ as

\begin{equation}
F_{\text{sat}} = 5 \phi \rho c^3_{s}
\end{equation}
where we have chosen $\phi = 0.3$ as recommended by \cite{BalbusMcKee1982}.  The second term $\bmath{F}_{\text{class}}$ is the classical heat flux that depends on the temperature gradients and magnetic fields. We compute the heat flux in its anisotrpic form using the Spitzer formula \citep{Spitzer1962,Braginskii1965}: 

\begin{equation}
  \bmath{F}_{\text{class}} = \kappa_{\parallel} \bmath{b} ( \bmath{b} \cdot
  \nab T ) + \kappa_{\perp} [ \nab T - \bmath{b} ( \bmath{b}
  \cdot \nab T ) ]
\end{equation}
where $\bmath{b} = \bmath{B}/ | \bmath{B}|$ is the direction of the magnetic fields. The coefficients $\kappa_{\parallel}$ and $\kappa_{\perp}$ are the thermal conductivities parallel and perpendicular to the fields \citep{Spitzer1962}. For completeness, the parallel conductivity is 

\begin{equation} \kappa_{\parallel}  = \varepsilon  \times 20 \left(\frac{2}{\pi}\right)^{3/2}   \frac{(k_B T)^{5/2} k_B}{m_e^{1/2} e^4 \ln \Psi} \end{equation}
where $\varepsilon$ is a coefficient of order unity ($\approx 0.4$) that reduces the Spitzer conductivity, $\kappa_{\parallel}  = \varepsilon \kappa_{\rm SP}$, to an effective conductivity due to thermoelectric effects. In addition, $\ln \Psi$ is the coulomb logarithm: 
\begin{equation}  \ln \Psi = 29.7 + \ln \left(  \frac{T / 10^6 \rm{K}}{\sqrt{n/\rm{cm^{-3}}}} \right) \end{equation} 
and the perpendicular conductivity has a smaller magnitude, and is defined as 

\begin{equation} \kappa_{\perp} = \frac{8 (\pi m_H k_B)^{1/2} n_H^2 e^2 c^2 \ln \Psi }{3 | \bmath{B}|^2 T^{1/2}} \end{equation}

\begin{footnotesize}
\begin{table}
\begin{center}
 \footnotesize
\centering
\begin{minipage}{160mm}
   \caption{Simulation suite}\label{tab:clsims}
   \begin{tabular}{@{}lcccc@{}}
     \hline
     \hline
	Simulations 	&  cooling	 	&  conduction  & magnetic field  & 			$\Delta x_{\rm res}$	\\
			     	 &   			&			 & $\beta = P_{\rm th} / P_B$ 	& (pc)	\\ 
     \hline
     $\beta_{\rm{ad},\infty}$ 		&  		  	& 			& $\infty$ 		& 0.37		 \\
     \hline
     $\beta_{\rm{fid},\infty}$ 		&  \checkmark  	& 			& $\infty$ 		& 0.37		 \\
     $\beta_{\infty}$ 				&  \checkmark  	& 			& $\infty$ 		& 0.73		 \\
     $\beta_{1000}$ 				&  \checkmark  & 		 	& 1000		& 0.73		 \\
     $\beta_{100}$ 				&  \checkmark  & 		 	& 100		& 0.73		 \\
     $\beta_{10}$ 				&  \checkmark  &  			& 10			& 0.73		 \\
 \hline
     $\beta_{c,\infty}$ &  \checkmark  		& \checkmark		& $\infty$			& 0.37		 \\
     $\beta_{c,100}$ 		&  \checkmark 	& \checkmark 		& 100			& 1.46		 \\
     $\beta_{c,10}$ &  \checkmark  		& \checkmark 		& 10				& 1.46		 \\
     $\beta_{c,1}$ &  \checkmark  		& \checkmark 		& 1				& 1.46		 \\
     \hline    
\multicolumn{5}{l}{All simulations have the same initial condition with hot bath } \\
\multicolumn{5}{l}{temperature at $T_{h} = 10^7$ K. The initial cloud is set up with $T\approx $} \\
\multicolumn{5}{l}{ $10^{4.3} - 10^{6.5} $ K. Magnetic fields are initialized parallel to the hot } \\
\multicolumn{5}{l}{  winds with velocity $v_h \approx 1.4 c_s$. }

   \label{line_table}
 \end{tabular}
\end{minipage}
\end{center}

\end{table}
\end{footnotesize}

To study the resulting sizes of cold clumps in a more realistic setting, we are interested in a region with varying density structure instead of a single spherical cold cloud.  To do so, we first initialize the entire simulation domain in pressure equilibrium at the hot gas density and temperature. We then add density perturbations (or equivalently entropy perturbations) that follow a random Gaussian field modulated by a power spectrum $P(k)\propto k^{-3}$. We then truncate the highest density region following \cite{McCourt2018} to avoid seeding of small-scale structures due to rapid cooling via its sensitive dependence with density $\dot{e} \propto n^2$. With small cold clumps emerging from thermal instabilities in many similar studies \citep[e.g.,][]{AuditHennebelle2005, Cooper2009,McCourt2018}, we do not expect our results to be strongly dependent on the details of the initial conditions. 

To embed a fractal cloud in the hot bath, we select a spherical region of the perturbed density field with a radius of 150 pc tapered by a Gaussian (Figure \ref{fig:IC}).  This fractal cloud is initially warm with a range of temperature from $\sim 10^{4.3} - 10^{6.5}$ K with a median temperature at $\sim 10^5 $K. We have also adopted a metallicity of the entire simulation domain at $Z = 0.3 Z_{\odot}$, which provides a maximum cooling rate near the median temperature. The hot wind is initialized with velocity $v_h \approx 1.4 \,c_s$, where $c_s$ is the sound speed of the ambient gas at $10^{7}$ K.   The high-temperature hot bath at $10^{7}$ K cools inefficiently at its density $n_{\rm H} = 10^{-4} \rm{cm^{-3}}$, which allows us to separate the cooling timescale of the targeted warm gas that will undergo thermal instability.  A variation of the hot bath temperature at $10^6$ K does not qualitatively change our results. 

Table \ref{tab:clsims} lists the suite of simulations in this study.  All of the simulations have the same initial condition as described above besides the variations of the input physical processes, such as cooling, thermal conduction, and magnetic fields. All simulations are labeled with $\beta = P_{\rm th}/P_{\rm B}$ as a parameterization for the initial strength of the magnetic field, ranging from $\beta = \infty$ (no magnetic fields) to $\beta = 1$. For a hot bath of $10^{7}$ K,  $\beta = 1$ corresponds to a magnetic field strength of  $\sim 2 \mu G$, which is typical of observations in star formation galaxies \citep[see][]{Beck2015}.  The subscript ``ad" indicates an adiabatic simulation without cooling ($\beta_{\rm{ad},\infty}$); and the subscript $``c"$ indicates thermal conduction is included in the simulations. The initial strength of the magnetic field is also subscripted for convenience. The initial direction of the magnetic fields is parallel with the hot winds for all of the simulations.

\subsection{The evolution of cold gas in simulations }

\begin{figure*}
\begin{center}
\hspace{30mm}
\includegraphics[width=18 cm,height=17 cm]{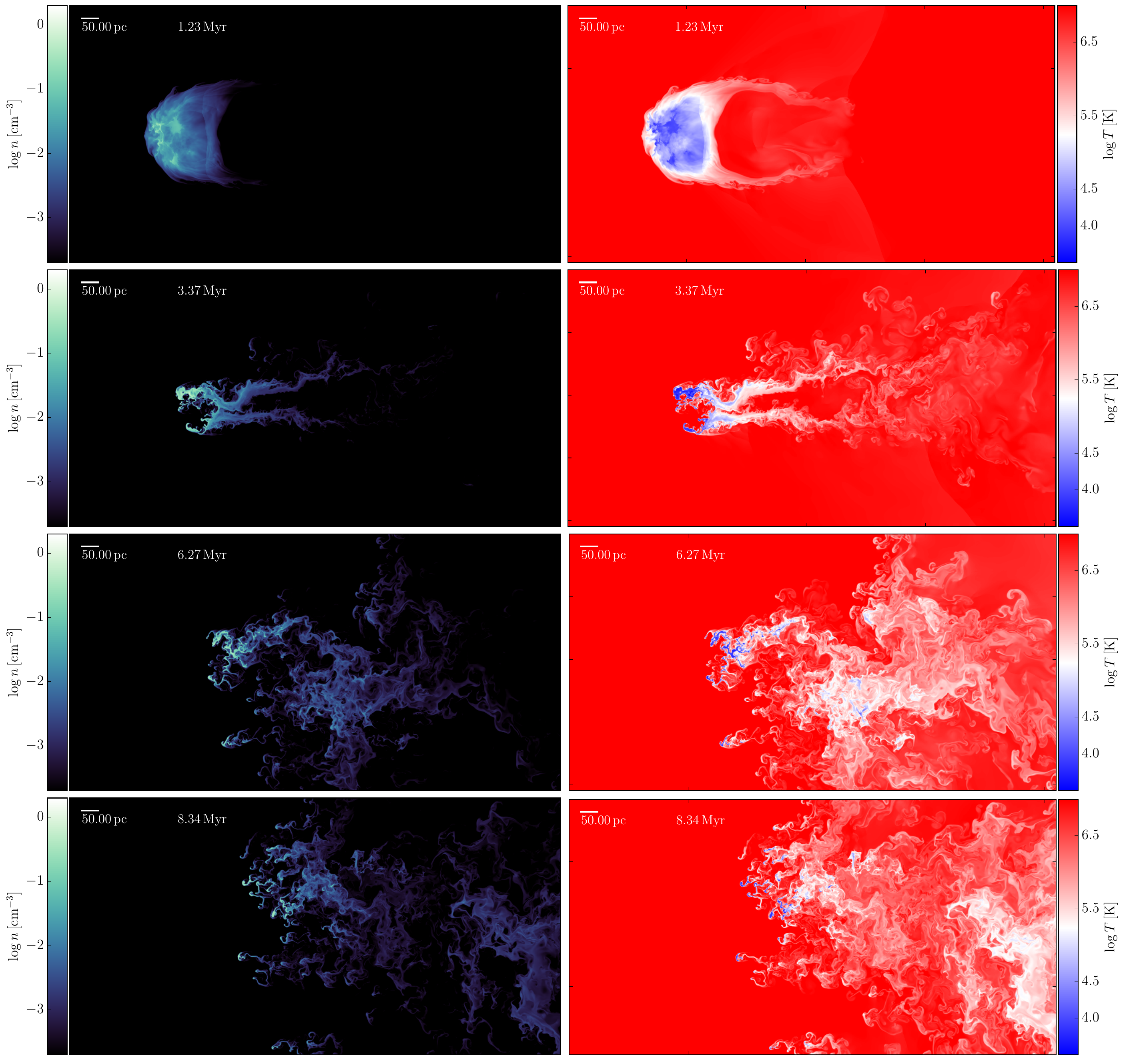}
\caption{The evolution of the density (left column) and the temperature (right column) in the fiducial simulation with cooling only ($\beta_{\rm fid} = \infty$). As shown, the initial warm gas cools quickly to $T\sim 10^{4}$K.  The cold gas then fragments into pc-size cloudlets through internal dynamics due to cooling and the interaction with the hot winds. \label{fig:fidsim}}
\end{center}
\end{figure*}

\begin{figure*}
\begin{center}
\includegraphics[width=17.5 cm,height=8 cm]{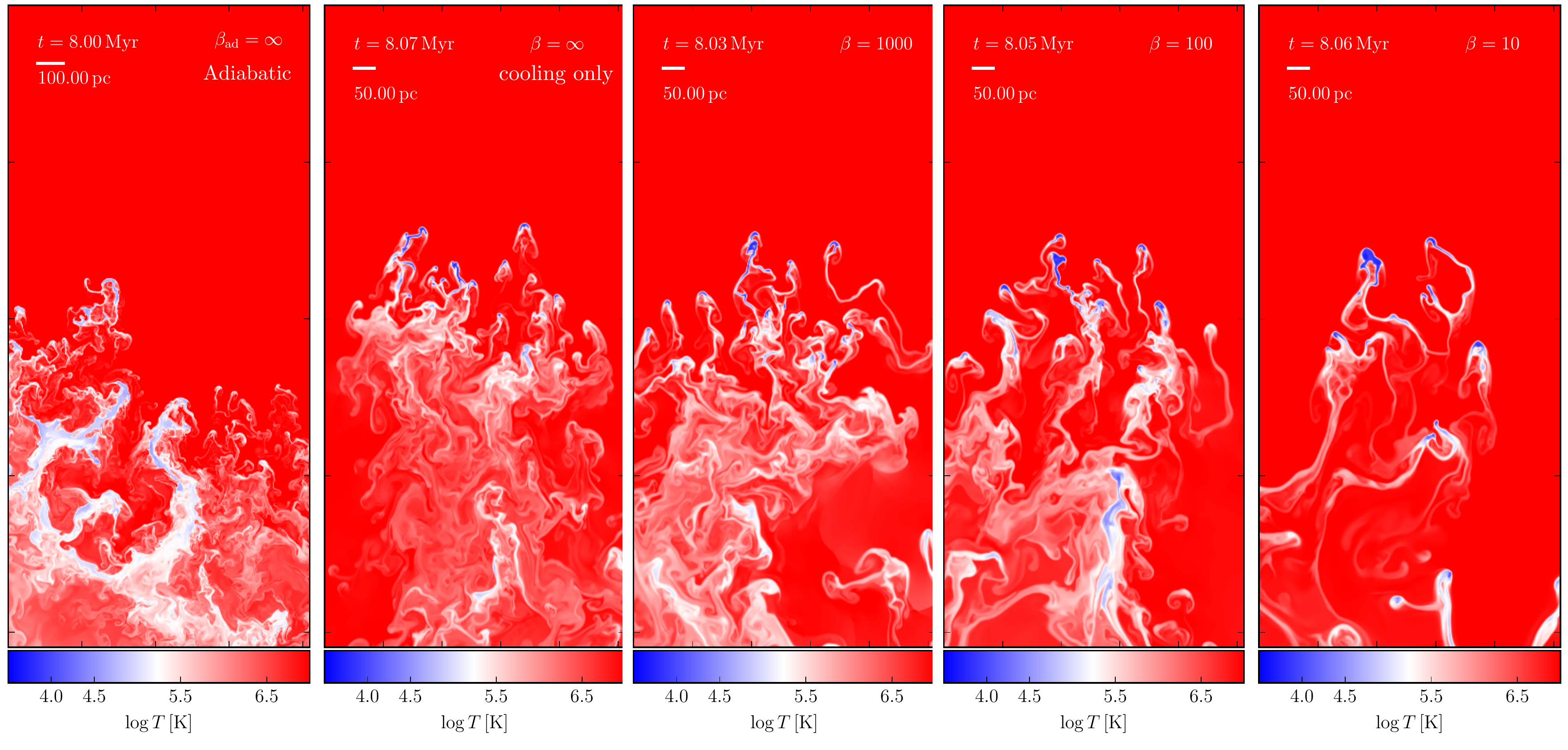}
\caption{Temperature maps of simulations with the same initial conditions except the strength of the initial magnetic fields ranging from  $\beta = 10 - \infty$. These simulations do not include thermal conduction. The left panel shows the adiabatic run where no cold clumps are present. This is in stark contrast to those with radiative cooling, where they produce similar cold clumps with $N_{\rm H} \sim 10^{17.5}\rm{cm^{-2}}$. Larger magnetic field strengths produce more filamentary structure (e.g., $\beta = 10$ in the right panel).  \label{fig:vertial_beta}}
\end{center}
\end{figure*}

\begin{figure*}
\begin{center}
\includegraphics[width=15 cm,height=21 cm]{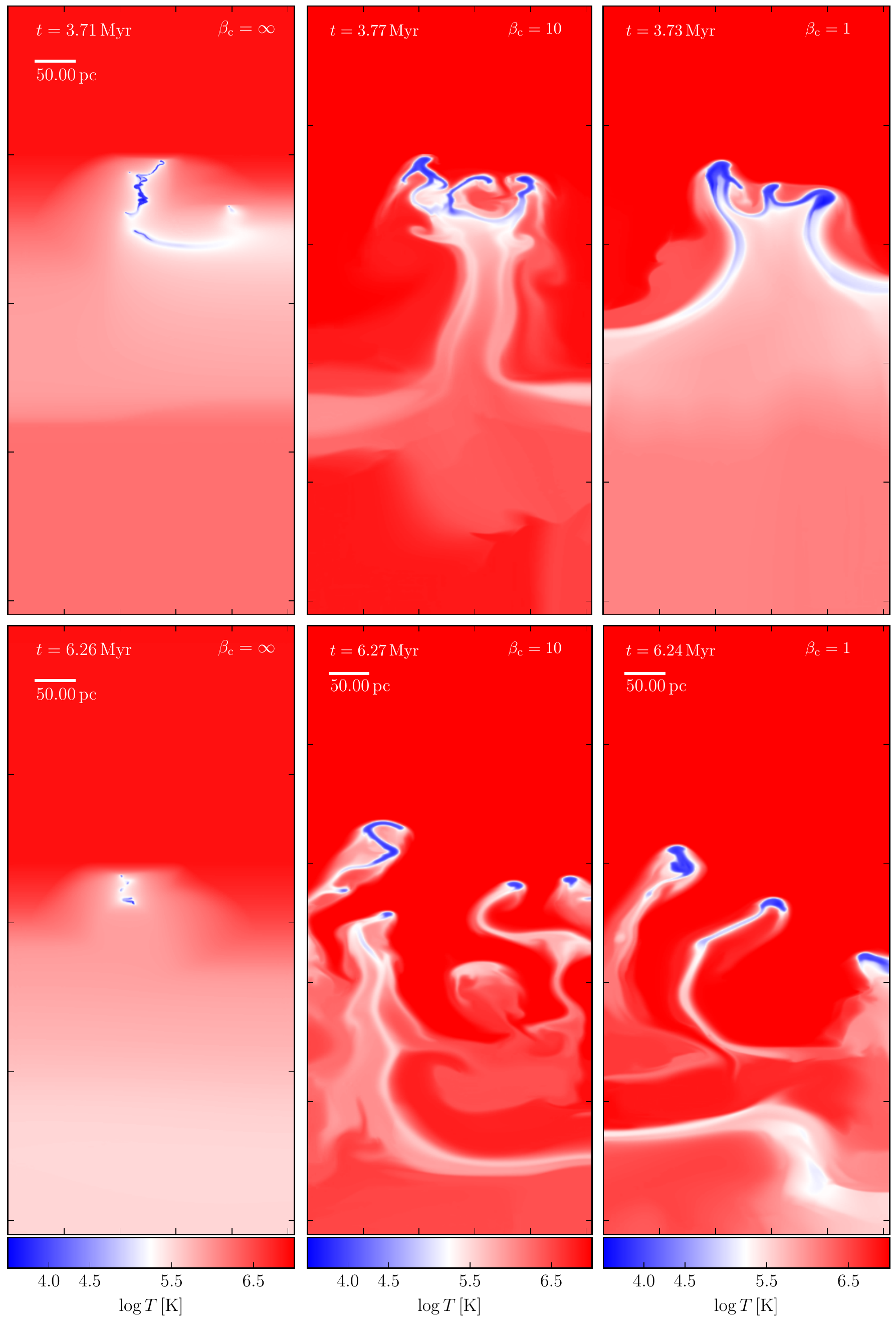}
\caption{Simulations with thermal conduction at two representative time steps. The left panel ($\beta_c = \infty$) shows a run with cooling and isotropic thermal conduction. The middle and right columns show the evolution with an increased magnetic field strengths, which prolong the survival of cold gas. \label{fig:vertial_betac}}
\end{center}
\end{figure*}

We begin this section with a qualitative description of the evolution of cold gas in our simulations. In particular, we discuss and compare with the relevant literature and focus on the destruction and fragmentation process and the resulting cloudlet sizes. 

\subsubsection{Adiabat: scale free evolution}

We begin with the adiabatic simulation ($\beta_{\rm ad, \infty}$).  Early numerical studies of the shock-cloud interactions have mainly focused on the competition of acceleration and destruction of spherical clouds.  It is well known that the evolution is self-similar in purely hydrodynamic systems where the problem depends only on the density contrast after scaling out the Mach number \citep[e.g., ][]{Klein1994, Nakamura2006}.  It is also well known that the Kelvin-Helmholtz and Rayleigh-Taylor instabilities disrupt the cloud in just a few cloud crushing times, where it is defined as:

\begin{equation} t_{\rm cc}  = \chi^{1/2}\frac{r_{\rm cl}}{v_h} = \left(\frac{n_{\rm cl}}{n_h} \right)^{1/2} \frac{r_{\rm cl}}{v_h}\end{equation}
where $n_h$ and $v_h$ is the density and velocity of the hot winds.  The adiabatic simulation can be thought as a baseline comparison for our study of the mist-like gas.  From Figure \ref{fig:vertial_beta}, it is clear that the purely hydrodynamical case does not exhibit any characteristic cold and dense cloudlets compared to all the other simulations that include radiative cooling. 

\subsubsection{Radiative cooling: cold clumps from thermal instability }

The effects of radiative cooling has also been analyzed in many papers in the context of entrainment of cold gas by hot winds \citep[e.g.,][]{Mellema2002, Melioli2005, Cooper2008,ScannapiecoBruggen2015, Schneider2018} and formation of multiphase ISM via thermal instabilities \citep[e.g.,][]{KritsukNorman2002, AuditHennebelle2005}. 

As radiative cooling is introduced to the simulations, it sets a qualitative different path for the evolution of the fractal cloud in our simulations. Most notably, a large fraction of cold gas is condensed out of the warm medium very quickly before the shock has a chance to break apart and penetrate through the more diffuse warm/hot gas (top left panel in Figure \ref{fig:fidsim}).  After some time, roughly two large clumps of cold gas formed from the initial density perturbations. Meanwhile, a long tail of warm gas follows as the gas is stripped by the hot winds.   Eventually, a combination of internal dynamics due to the lost of thermal pressure and the hot winds are able to break apart the cold clumps through the path of least resistance, which is regions of slightly lower density \citep[see also][]{Cooper2009, SchneiderRobertson2017}.  Similar to results in our paper, fragmentation via thermal instabilities allow small clumps of cold gas to survive longer \citep{Mellema2002, Fragile2004, McCourt2018}, though they eventually mix with the ambient hot gas. We find similar conclusion as \cite{Cooper2009}, where the number of cold clumps increases with increasing resolution of their simulations. 

\subsubsection{Cooling and thermal conduction: elongation and evaporation}

The evolution changes very drastically as isotropic thermal conduction is added to the simulation ($\beta_{c,\infty}$; see left column in Figure \ref{fig:vertial_betac}). Initially, the warm gas cools as in the simulations with cooling only. At the same time, a conduction front propagates from outside in due to temperature gradients with the hot bath.  At this point, most of the density and temperature fluctuations are erased except the densest cold gas.  Although our initial condition of the cloud morphology is drastically different (fractal cloud), the morphology of cold gas at late times is similar to the findings in \cite{BruggenScannapieco2016}. That is,  the cold gas is shaped into an elongated filament by a combination of thermal conduction and the hot winds (see Figure \ref{fig:vertial_betac}).  This is because any cold gas structures perpendicular to winds are most susceptible to complete destruction. Thus, the cold gas that is shielded from behind survives for the longest. Naturally, this leads to a minimization of the cross-section area of cold gas with the hot winds, leaving an elongated column of materials at late times \citep[see also Figure 5 in ][]{BruggenScannapieco2016}.  This minimization of cross-sectional area and the high column of mass makes it more difficult to accelerate by ram pressure of the hot winds since the acceleration $a_{\rm ram} \propto \frac{A}{m_{\rm cl}}$.  Furthermore, this column of cold gas evaporates in just 7 Myr, much more quickly than its counterpart without conduction ($\gtrsim$ 10 Myr with cooling only). Ubiquitous detection of cold gas in galactic winds and halos suggests that thermal conduction may be suppressed. 

\subsubsection{Cooling, magnetic fields, and conduction: filamentary cold gas}

It is natural to include magnetic field in this problem since it not only prolongs the lifetimes of cold gas to help solve the entrainment problem \citep[e.g.,][]{MacLow1994, McCourt2015, Zhang2017}, it also shapes the morphology and size of the resulting cold gas from thermal instabilities \citep[e.g.,][]{Shin2008}.  In this study, we run a set of simulations with a range of initial magnetic field strength. The weak magnetic field simulation ($\beta = 1000$) essentially show the same number of fragments and very similar evolution as the simulation with cooling only. Very quickly, the initial fields evolve and are draped around individual cloudlets as they form. The hot winds and draping of magnetic fields lead to a substantial increase in magnetic energy density around the cold gas.  At late times, the interface of cold and hot gas is stabilized by strong magnetic fields, which suppresses fragmentation and mixing \citep[e.g.,][]{Shin2008}. This also provides some protection for the cold gas against thermal conduction and preserves large temperature gradients.  The cold gas generally survives for a longer period of time compared to simulations without magnetic fields.  Investigating the competition between the destruction by hot winds and conduction, and the protection by magnetic fields is out of the scope of this paper and deserves its adequate attention; we will leave this to a future study.

As we increase the initial strength of the magnetic fields, its energy density around the cold gas increases large enough that they start to play a dynamical role \citep[e.g.,][]{DursiPfrommer2008}. As shown in Figure \ref{fig:vertial_beta} and Figure \ref{fig:vertial_betac}, the number of cloudlets decreases with increasing magnetic field strength since it becomes more difficult to break apart the cold gas with the protection of the fields.  The morphology becomes more and more filamentary. The tail of warm gas also becomes less turbulent and more coherent.

\begin{figure}
\begin{center}
\includegraphics[width=\columnwidth]{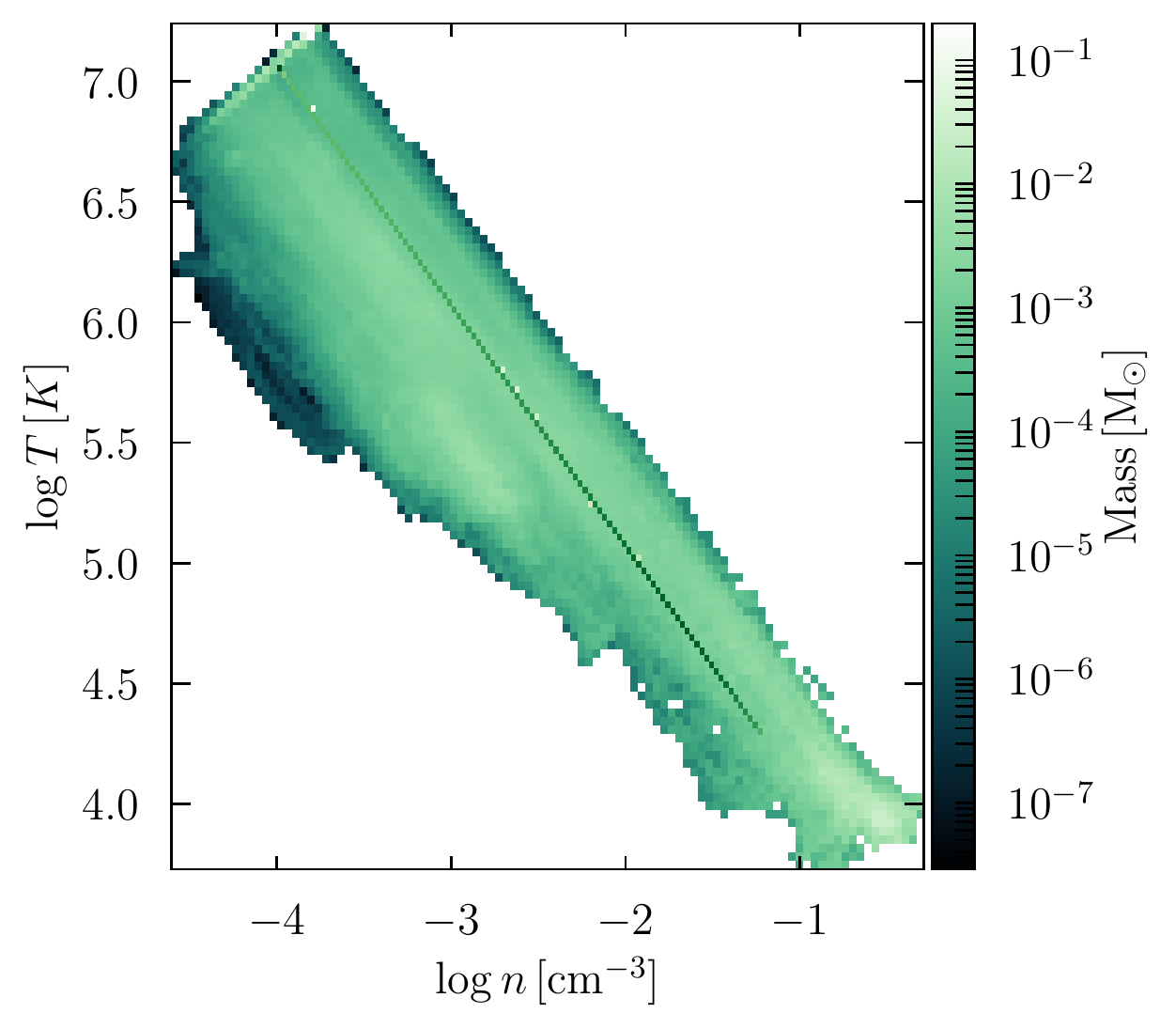}
\caption{Temperature and density distribution of all the computational cells of the cooling only simulation. The dark green line shows the initial distribution where the simulation is initialized in pressure equilibrium. The broader distribution shows the gas maintains approximate pressure equilibrium with the ambient hot gas. \label{fig:pequil}}
\end{center}
\end{figure}

\subsection{Column density of cloudlet comparisons in simulations}


\begin{figure}
\begin{center}
\includegraphics[width=\columnwidth]{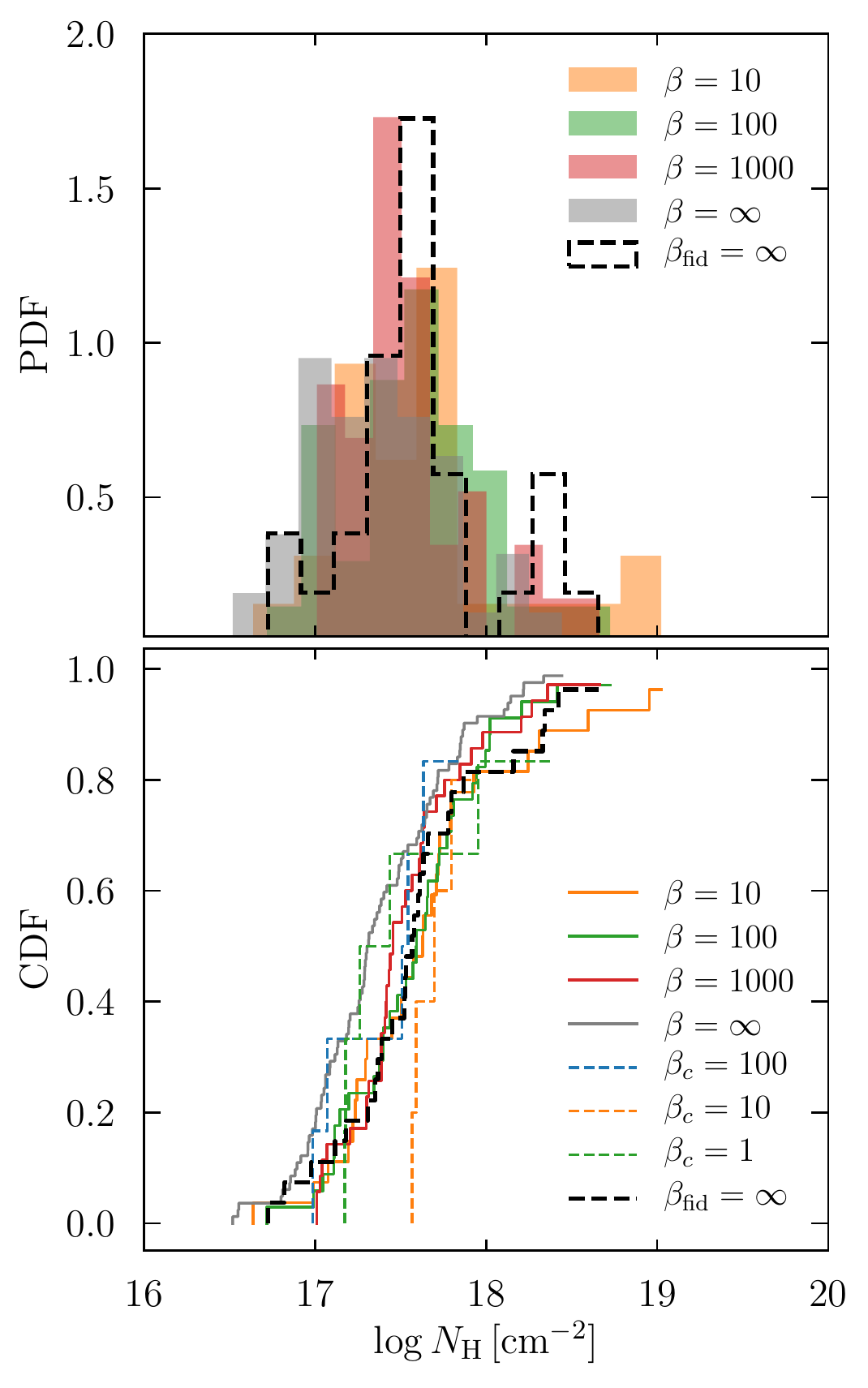}
\caption{The top panel shows column densities distribution for simulations without thermal conduction.  Simulations with conduction are better shown without binning in the cumulative distribution in the bottom panel, since there is only a small number of cloudlets. As shown, the medians of these distributions are all located near $N_{\rm H} \approx 10^{17.5}\rm{cm^{-2}}$, as predicted by the model. Simulations with large initial magnetic field strengths (e.g., $\beta_c = 1-10$) show the cloudlet column densities are slightly higher.  \label{fig:NHdist}}
\end{center}
\end{figure}

To directly verify the model ansatz that the size of cloudlet follows $r_{\rm cl} = \rm{min}(c_s t_{\rm cool})$, it requires high resolution simulations to resolve the gas clumps at a scale of $\sim 0.1 - 1$ pc. We expect decreasing clump sizes and increasing density with increasing spatial resolution if they are not converged in simulations.  In an ideal situation, one should run simulations with varying resolution and prove convergence of cloudlet sizes. Given the constraints of computational expense, it is difficult to do so directly. Instead, we consider the column density per cloudlet, $N_{\rm H} \sim 2 r_{\rm cl} n_{\rm cl}$,  since the size and the density goes in the opposite direction with a change of the resolution.  This is essentially equivalent to a statement of mass conservation. That is, the column density should also be insensitive to the resolutions in our simulations. In addition, the model predicts a characteristic value of column density at $N_{\rm H}\sim 10^{17 - 17.5}\, \rm{cm^{-2}}$, which is insensitive to the background pressure.  Therefore, we will measure column density per cloudlet in all of our simulations.  Nevertheless, we run a convergence study of column density of cloudlets using cooling only simulations with two resolutions ($\Delta x = 0.37, 0.74$ pc) given the computational power constraints we have. We find that the column density does not change dramatically (see comparisons in Figure \ref{fig:NHdist}). While one can argue that convergence has not been satisfactorily demonstrated,  for the purposes of our discussion of pc-scale cold gas in many kpc-scale gaseous halo, these differences are unimportant.

To measure the column density of a cloudlet, we first extract regions of cold gas using a density threshold of $n \ge 0.01 \,\rm{cm^{-3}}$. We find that the exact value of this density threshold does not change the result very much by varying its value from $10^{-3} - 10^{-2} \,\rm{cm^{-3}}$. After locating the cloudlets, it is easy to measure the column density by summing density over a line of sight. However, the morphology of cold gas is not always close to spherical. In fact, the cold gas is more filamentary in the presence of magnetic fields. The column density thus depends on the orientation of the lines of sight. Therefore, we consider the median of the column densities distribution by sampling all orientation as a representative value for a cloudlet. 

Figure \ref{fig:NHdist} shows the distribution of column densities of cloudlets from all of the simulations listed in Table \ref{tab:clsims}.  The top panel shows the probability density function (PDF) of simulations with cooling and magnetic fields.  Simulations with conduction are not included in the PDF since there are only a small number of cloudlets ($\sim 3-5$). Instead, it is better shown in the cumulative distribution function (CDF) without the need of binning. The CDF also includes the other simulations without conduction for comparison.  As shown, all column densities are distributed around $N_{\rm H} \sim 10^{17.5}\,\rm{cm^{-2}}$. As mentioned above, strong magnetic fields alter the dynamics of cold gas. We see hints in our simulations that stronger fields lead to bigger cloudlets due to magnetic tension. This is reflected in the small shifts towards larger cloudlet column densities in simulations with large initial magnetic fields (i.e $\beta_c =1$, see Figure \ref{fig:NHdist}).  However, as \cite{DursiPfrommer2008} pointed out, the dynamical process results from magnetic draping is an inherently three-dimensional problem. A detailed exploration of the dynamical importance of magnetic fields in this scale requires a parameter study in three-dimensional simulations in the future.

\section{Implications of cloud sizes in the Circumgalactic context}
\label{sec:rclcgm}

\begin{figure*}
\begin{center}
\includegraphics[scale = 0.7]{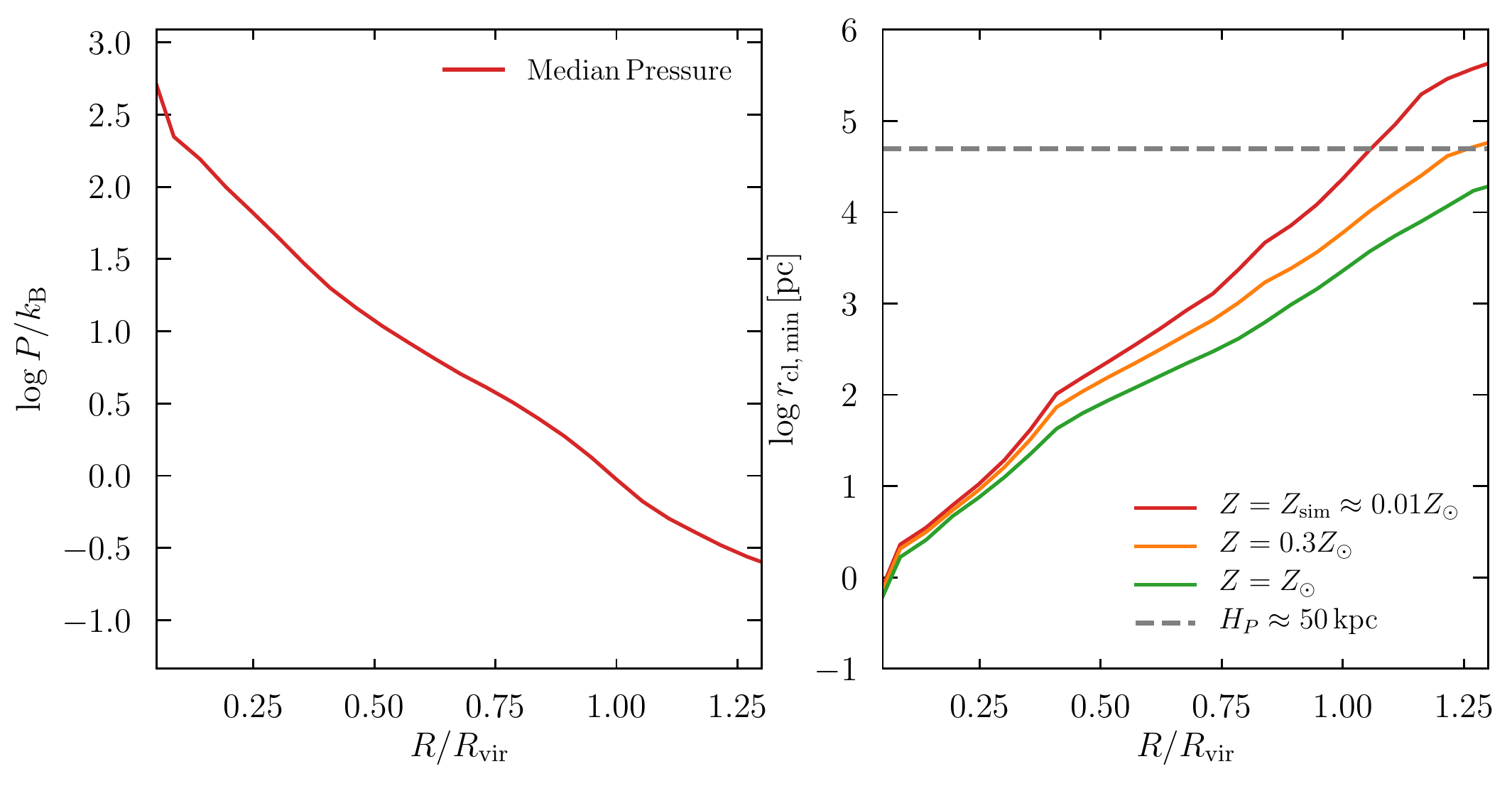}
\caption{Left panel: simulation median pressure profile of $L_*$ like galaxy at redshift $z = 0$  with $M_{\rm vir} = 10^{12} M_{\odot}$\citep{AgertzKravtsov2015,Liang2016}. Right panel: corresponding allowed cloudlet sizes, $r_{\rm cl,min} = \rm{min}(c_s t_{\rm cool})$, as a function of radius based on the pressure profile. Different color curves show the dependence of metallicity. The higher the metallicity, the more efficient the gas can cool, which leads to smaller cloudlet sizes. Note that the sharp boundary of detections of low and intermediate ions at $0.5-0.7 R_{\rm vir}$ \citep{LiangChen2014} corresponds to similar radii where small cloud sizes are small, in the order of 100 pc or less. In addition, the pressure profile on the left can be approximated by an exponential profile, $P/k_B\propto \exp[-(r/H_P)]$, where $r = R/R_{\rm vir}$ and $H_P \approx 50$ kpc is pressure scaleheight.  We expect the assumption of pressure equilibrium to derive the cloudlet size breaks down roughly at the size of the pressure scaleheight. \label{fig:rclpro}}
\end{center}
\end{figure*}

We now turn our attention to discuss the implications of cloudlet sizes on the larger context, namely the global picture of the gaseous halos around galaxies.

\subsection{The circumgalactic mist in galactic halos}


To paint a physical picture of the circumgalactic mist, let us consider a parcel of hot halo gas with size $L$, density $n_h$, and temperature $T_h$. We are particularly interested in the multiphase nature of the gas as a function of these background properties. Let us assume that this parcel of hot gas is located at some galactocentric radius $R$, where its cooling time is short compared to the global free-fall time:

\begin{equation} t_{\rm cool} / t_{\rm ff} < \xi \end{equation}
where $\xi$ is a parameter that controls whether the parcel of gas can cool and condense.  It has been shown halo gas can undergo thermal instability and precipitate at the mass scale of galaxy clusters if $\xi \lesssim 10$. \citep[e.g.,][]{McCourt2012,Voit2015b}. The cycles of condensation have been shown to be promoted by active galactic nuclei (AGN) feedback \citep[e.g.,][]{Gaspari2013b, Li2015, Meece2015}. While there exists indirect evidence for this same precipitation limit for lower mass halos \citep{Voit2015a}, we nevertheless will treat this as a free parameter.  Given a fixed $\xi$, we can find that a new hot gas density $n_{h,s}$, such that it is thermally stable. This can be achieved with the reverse condition, $t_{\rm cool} / t_{\rm ff} \ge \xi$, which implies\footnote{We have assumed equilibrium between electrons and the ions, $n_e = n_i = n_{h,s}$.},

\begin{equation} n_{h,s} \le \frac{3k_{\rm B} T_h}{\xi \Lambda t_{\rm ff}}  \end{equation}
Thus, the gas left over from condensation has a long cooling time due to its relatively low density. Again, the global balance of energy accounting can be maintained and supported by feedback processes and photoionization.  While we are interested in the cold gas out of the condensation, this hot gas density profile has its own implication. In fact, \cite{Voit2018} derived a hot density profile using $\xi \ge 10$ and showed X-ray luminosity $L_X$ of halos obeys this precipitation limit across a large range of mass scales, from clusters, galaxies groups to massive ellipticals \citep{Cavagnolo2009, Sun2009, Werner2012, Werner2014}.

Let us now consider the cooling phase of this parcel of gas. First, we assume some density perturbations $\delta n / n_h$ with scale less than $L$, such that a fraction of this parcel of gas can undergo thermal instability with fragmentation and contraction.  At this stage, we are not concerned with how this portion of the mass distributes itself over a range of temperature over time. The mass of this cooling phase simply is the difference between the original total mass and the leftover stable hot gas:

\begin{equation} M_{\rm cooling} = L^3 \mu m_H (n_h - n_{h,s}) \end{equation}
where $\mu$ and $m_H$ is the mean molecular weight and mass of the hydrogen atom, respectively. 
Note that $M_{\rm cooling} = 0$ if the gas is already stable with $n_h - n_{h,s} \le 0$, which is the case at large radii when $t_{\rm cool} > 10 t_{\rm ff}$ \citep[e.g.,][]{Voit2015a}. The available cooling materials are limited in the inner regime, and provide a natural explanation for decreasing covering fraction of cold gas with increasing radius \citep{LiangChen2014}. However, this does not mean cloudlets can exist as long as $t_{\rm cool}/t_{\rm ff} < 10$ because the background pressure is important as well. 

As in \cite{Spitzer1956}, we assume the cold cloudlets are pressure confined by the hot gas. 
One can then calculate the mass of an individual cloudlet, since $n_{\rm cf,f}$ and $r_{\rm cl}$ are both functions of the ambient pressure (see section \ref{sec:rclmodel}):

\begin{equation} m_{\rm cl} = \frac{4\pi}{3} r_{\rm cl,f}^3  \mu m_H  n_{\rm cl,f} \end{equation}
 In the inner region of the CGM, a large fraction of the cooling mass $M_{\rm cooling}$ would have reached $T_{\rm cl,f} \sim 10^4 $K very quickly in the form of a cloudlet (section \ref{sec:rclmodel}). In this simplified picture of spherical cloudlets, we can also estimate the total number, $\mathcal{N_{\rm cl}} $, within this region of size $L$:

\begin{equation}  \mathcal{N_{\rm cl}} = \frac{M_{\rm cooling}}{m_{\rm cl}} =  \frac{3}{4 \pi} \left(\frac{n_h - n_{h,s}}{n_{\rm cl,f}}\right) \left(\frac{L}{r_{\rm cl}}\right)^3 \end{equation}
 It also useful to consider the number of cloudlets per volume, which is independent of size of the domain $L$,

\begin{equation}  \eta_{\rm cl} = \frac{\mathcal{N_{\rm cl}}}{L^3} =  \frac{3}{4 \pi} \left(\frac{n_h - n_{h,s}}{n_{\rm cl,f}}\right) \left(\frac{1}{r_{\rm cl}}\right)^3 \label{eqn:rcldensity}\end{equation}
From this,  we can also calculate the average spacing between cloudlets: 

\begin{equation} \langle d_{\rm cl} \rangle  = \eta_{\rm cl}^{-1/3} \approx \left(\frac{n_{\rm cl,f}}{n_h - n_{h,s}}\right)^{1/3} r_{\rm cl} \end{equation}
The number of cloudlets depends strongly on the size of the individual cloudlet. For the inner region of the CGM, as shown in Figure \ref{fig:rclpro}, the size of the cloudlets is in the order of $r_{\rm cl} \sim 0.1$ to a few parsec. Given a region with $L = 1$ kpc within $\sim 0.25 R_{\rm vir}$, the ratio $\left(\frac{L}{r_{\rm cl}}\right)^3  = 10^{9} - 10^{12}$. If we take a typical value of $n_h \sim 10^{-4} \rm{cm^{-3}}$ and $n_{\rm cl,f} \sim 10^{-1}\rm{cm^{-3}}$, the ratio of the densities is roughly $(n_h - n_{h,s})/n_{\rm cl,f} \sim n_h /n_{\rm cl,f} \sim 10^{-3}$.  With the assumption of spherical cloudlets, there exists at least in the order of $\sim 10^6- 10^9$ clumps  within a 1 kpc region! The average inter-cloudlet spacing is in the order of 30 cloudlet radii. In this picture, the cloudlets forms a kind of mist, resembling terrestrial clouds and mist.

\cite{McCourt2018} calculated the expected voluming filling and covering fractions of cloudlets (see their Figure 1). For completeness, we can compute these quantities now with an expression for the number density of cloudlets (Eq. \ref{eqn:rcldensity}). The volume filling factor is simply the ratio of the total volume occupied by the cloudlets over the volume of the region,

\begin{equation}  f_{\rm V} =  \mathcal{N}_{\rm cl} \left(\frac{r_{\rm cl}}{L}\right)^3 = \mathcal{\eta}_{\rm cl} r_{\rm cl}^3 \approx  \frac{n_h - n_{h,s}}{n_{\rm cl,f}} \label{eqn:fvol}\end{equation}
Therefore, the volume filling factor of the mist is actually quite small at $\sim 0.1\%$ of the total volume $L^3$ in the inner CGM.  However, this does not imply the covering fraction is small. To get an estimate, we can start by assuming cloudlets are spherical with a cross sectional area, $\sigma_{\rm cl} = \pi r_{\rm cl}^2$. If they do not shadow each other, the covering fraction integrated over some length scale, say $L$, is equivalent to the formula for optical depth: 

\begin{equation} f_{\rm A} = \frac{\sigma_{\rm tot}}{A} = \frac{\mathcal{N}_{\rm cl} \sigma_{\rm cl}}{A} =  \frac{\eta_{\rm cl} LA \sigma_{\rm cl}}{A} =\eta_{\rm cl} \sigma_{\rm cl} L\end{equation}
where $\rm{A}$ is some area of interest for calculating the covering fraction.  With the expression for cloudlet number density, $\eta_{\rm cl}$, the covering fraction is: 

\begin{equation} f_{\rm A} \approx \frac{n_h - n_{h,s}}{n_{\rm cl,f}} \frac{L}{r_{\rm cl}} \end{equation}

Therefore, the covering fraction can quickly reach unity for $r_{\rm cl} \sim 1 $ pc over a length scale of 1 kpc. This is in concordant with observations of Mg\,II absorbers close to the galaxies where the covering fraction is very high \citep{Nielsen2013, Chen2010, Huang2016}, even if the volume filling factor can be small \citep[e.g.,][]{Hennawi2015}. In fact, the large number of cloudlets may also affect the degree of resonant scattering with Ly$\alpha$ photons and have observable signatures in their emission spectra \citep[e.g.,][]{Gronke2017}. We also note the morphology of cold gas needs not be spherical. The assumption and simplification to sphericity provide a simple way to calculate a physical picture of the mist. In reality, the morphology can be more complicated. For example, the cold gas can be filamentary in the presence of magnetic fields as shown in this paper. This is especially true in the picture that magnetic field lines are draped around cold gas to prolong lifetimes of cold gas within galactic outflows and the inner halo. 

\subsection{Spatial extents of low, intermediate and high ions from the galactic profile of cloud sizes}

In \cite{Spitzer1956}, it was proposed that the observed cold gas \citep[Na\,I and Ca\,II; ][]{MunchZirin1961} must be pressured confined to explain their existence at large ($\sim$ few kpc) distances away from the galactic plane. Otherwise, the number density of such ions will decrease as they expand; the atoms will be ionized into a higher state. The absorption by these low ions will, therefore, be very slight.  It is natural to consider what happens over larger distances where the pressure profile does indeed drop exponentially.  The CGm model naturally predicts the change in density and sizes of resulting cloudlets from the pressure profile.  

The abundance of low, intermediate and high ions peak at a range of preferred density, or ionization parameter ($U$) given a fixed photoionization background.  In other words, there will be more ionizing photons per atom (higher $U$) at larger distances.  Therefore, if the gas is primarily photo-ionized, it will be elevated to higher and higher ionization states as the cloudlet size increases and its corresponding density drops \citep[see similar discussion in][]{Liang2016, Liang2017}. This in turns provides predictions for the spatial extent of commonly observed low, intermediate and high ions. 

The left panel of Figure \ref{fig:rclpro} shows the pressure profile of hot gas from a cosmological ``zoom-in" simulation of a Milky-Way size galaxy at $z = 0$ \citep{AgertzKravtsov2015, Liang2016}. The halo mass at this time is $M_h \approx 10^{12} M_{\odot}$ with a virial radius of $R_{\rm vir} \approx 260$kpc. Given the description in section \ref{sec:rclmodel}, we can calculate the corresponding \textit{allowed} cloudlet sizes as a function of distance given the ambient pressure, as seen in the right panel of Figure \ref{fig:rclpro}.  Since the final cloudlet size depends also on how effectively the gas can cool through the cooling function $\Lambda$, we show different color lines bracketing a range of sizes from metallicity effects, with $Z/Z_{\odot} = $0.01, 0.3 and 1. While in reality there exists pressure variation within the halo, the lines in the right panel of Figure \ref{fig:rclpro} represents the cloud sizes from the median pressure at a given radius. In addition, the pressure profile can be approximated by an exponential profile, with $P/k_B\propto \exp[-(r/H_P)]$, where $r = R/R_{\rm vir}$ and $H_P \approx 50$ kpc as the pressure scaleheight. This implies that the pressure equilibrium assumption, at the very least, breaks down above this scale, which puts a limit on the applicability of the model. This occurs near the virial radius of the halo for this particular $L_*$ galaxy. 

The cloudlet sizes $r_{\rm cl}$ and the densities ($n_{\rm cl}$, not shown) span many orders of magnitude within the halo. It is interesting to point out the characteristic sizes at a few different radii. First, the cloudlets are tiny with $\lesssim 1$ pc at small distances where the background pressure is the highest \citep[see also fragment sizes in][]{AuditHennebelle2005}. Second, observed cloud sizes derived from cold/warm absorbing gas generally fall between tens to hundreds of parsecs \citep[e.g.,][]{Rauch1999, ProchaskaHennawi2009, Crighton2015,  Stern2016, LanFukugita2017}. The $\sim$100 pc scale of cloudlets corresponds to distances at $0.5 R_{\rm vir}$ or $\sim$130 kpc. This is in concordant with the observed spatial extent of Mg\,II, Si\,II and C\,II \citep{Chen2010, Werk2013, Nielsen2013, LiangChen2014, Huang2016}. In other words, not everywhere within the galactic halo can the mist exist (i.e., small, cold, and dense cloudlets).  

Similarly, intermediate ions that trace warm gas preferred a range of slightly lower density and larger cloud sizes. This allows ions like C\,IV, and Si\,IV to be present at larger distances from the host galaxy. Indeed, these have been observed up to $\sim 0.7R_{\rm vir}$ \citep{LiangChen2014}. These authors also showed differential covering fractions from low to high ionization state for a fixed specie (e.g., Si\,II, Si\,III and S\,IV). 

Beyond $\sim 0.7R_{\rm vir}$, the cloudlet sizes increase to many kiloparsecs, approaching the size of the hot halo. In this limit, only high ions that trace the hot gas can survive. Photoionization may not play a big role for high ions since the ionization energy is high (e.g., $E_{\rm ion} = 138$ eV for O\,VI). Instead, O\,VI tracing this component of the warm/hot gas is most likely collisionally ionized. However, the ionization mechanism of O\,VI is being debated from both observational and theoretical perspectives \citep[e.g.,][]{ThomChen2008, Tripp2008, Werk2016, Faerman2017, MathewsProchaska2017, Liang2017, Stern2018, Nelson2018}. Nevertheless, QSOs absorption-line observations show that O\,VI is detected up to $R_{\rm vir}$ \citep[e.g.,][]{Johnson2015}.  The abundance of O\,VI observed in the outskirt implies that the gas somehow sustains hot despite its high cooling rate at ~$10^{5.5}$ K. The question of how the hot gas remains in this state has to do with the global thermal balance from various heating mechanisms. As mentioned in the introduction, the subject of feedback processes and its regulation of star formation and the hot halo are being investigated \citep[e.g.,][]{Oppenheimer2016, Fielding2017}. 

\section{Discussions and Conclusions}

Throughout this paper, we have loosely used the terms, circumgalactic mist and cloudlet, to paint a general picture of cold gas in galactic outflows and the inner CGM owing to their similarities with condensation in thermal instabilities.  This analogy with terrestrial clouds goes back many decades in the literature. In fact, condensation of cold gas has been explicitly compared with the van der Waals gas \citep{ZeldovichPikelner1969}.  However, it is worth pointing out that there are no natural boundaries in which one can easily define the discreteness of a cloudlet. While the underlying language that describes the physics of the CGM is mathematical (e.g., hydrodynamics and turbulence), the choice of words is only used to paraphrase the general properties of the cold gas. In reality, the model describes a region of the CGM with local density variations at which thermal instability can be excited within this compressible turbulence. Afterward, the competing timescales imprint a spatial scale ($c_s t_{\rm cool}$) for cold gas.  We show in this paper thermal instability can be triggered by a combination of supersonic galactic winds and initial density perturbations in a hot bath. Just as terrestrial clouds and mist can form when a region of hot and humid air is hit with a cold front, an episode of cold gas can form within a compressible turbulence induced by supersonic hot winds.

The abundant observations of cold gas in the inner region of the CGM suggest that it is a dominant evolutionary phase of the galaxies. In fact, the cold phase of the CGM is not only ubiquitous across cosmic times \citep[e.g.,][]{Steidel2010, Chen2012, Werk2013,LiangChen2014}, but also across galaxy types \citep[e.g.,][]{Prochaska2014, Bordoloi2014, Huang2016, Heckman2017}.  The origin of this dynamic yet steady multiphase CGM is not yet fully understood. There are competing scenarios for the origins and implications of cold gas. For example, can cold gas be directly launched from the ISM into the halo by entrainment with the help of magnetic fields \citep{McCourt2015, Zhang2017}? Fragmentation may additionally prolong the lifetime of cold gas due to inefficient mixing and allow cloudlets to be more easily entrained. At the same time, cold gas can be formed directly within the galactic winds and halos via thermal instability as presented in this paper. A third option for the formation of cold gas can be mass-loaded winds through adiabatic cooling \citep[see also][]{Thompson2016, Schneider2018}. Thus, it is worth exploring any observables, such as kinematics signatures, that could distinguish the different scenarios. 

While the simulations in this paper represent an episode of the birth and death of cold gas, the analytic CGm model describes, in a sense, a steady state of the multiphase CGM.  The spatial extent depends mainly on the ambient pressure profile. The abundance of cold halo gas can fluctuate over many episodes of this process; it can also depend on other processes such as AGN feedback in more massive systems, which fluctuates in the order of many millions of years \citep[e.g.,][]{Li2015}. However, comparison between observations of low ($z \approx 0$) and high redshifts ($z \approx 2$) show remarkable similarities over a much longer timescale \citep[e.g.,][]{Steidel2010, Chen2012,Prochaska2013, LiangChen2014}. Galaxy simulations that reproduced the observed low-$z$ CGM also suggest that the CGM evolves very little over cosmic time once it is properly normalized in units of the dark matter halo scale radius $r_s$\citep{Liang2016}. The unchanging nature of the CGM over cosmic time is somewhat remarkable and puzzling because it implies that the spatial extent of cold gas depends little on the star formation rates of galaxies where it is an order of magnitude higher in the cosmic high noon at $z \approx 2$ compared to today.  To understand this, we must model self-consistently in our galaxy simulations the processes of star formation, feedback and thermal instabilities that lead to a multiphase CGM over cosmic time. 

It is clear that the current generation of galaxy simulations do not resolve the million of gas clumps within $ \lesssim 0.5 R_{\rm vir}$ with a typical resolution of $\Delta x_{\rm res} \sim 1$ kpc in the halo. This is because current state-of-the-art simulations focus on their adaptive refinement criteria at the highest density regime. The common underproduction of cold halo gas by cosmological galaxy simulations may be due to the poor resolution in the halo. The profiles of cloud sizes in Figure \ref{fig:rclpro} and Eq. \ref{eqn:rcldensity} suggest that future galaxy simulation should address the mismatch of cloud sizes and their spatial resolutions, in order to correctly predict the clumping factor and consequently cooling of the halo gas. It implies that the resolution in the halo within $\sim 0.5 R_{\rm vir}$ must be in the order of 100 pc or less to match the observed abundance of cold gas \citep[e.g.,][]{Chen2010, Nielsen2013, Werk2013, LiangChen2014, Huang2016}.  An alternative is to employ some types of subgrid models that incorporate the process of fragmentation and thermal instabilities, especially for galaxy simulations with large cosmological volume \citep[e.g.,][]{Vogelsberger2014, Schaye2015}.  Resolving this process may also introduce stronger coupling between the multiphase CGM and the central star-forming sites and lead to a better understanding of their symbiotic relationship.

In summary, we have presented a new multiphase model of the CGM, the circumgalactic mist. Our conclusions can be summarized as follows: 

\begin{enumerate}

    \item[\textbf{1.}]  We considered the process of fragmentation and isobaric condensation modes in thermal instability and showed that the resulting cloudlet sizes are very small ($\sim$pc) in the galactic hot halo environment. We provided a semi-analytical prescription to derive the cloudlet sizes entirely based on background hot halo properties (section \ref{sec:rclmodel}). 

    \item[\textbf{2.}]  Using a set of MHD simulations, we discuss the process of fragmentation and the birth of cold gas in general. We show that the resulting cold gas conforms to the characteristic column density of $N_{\rm H} \approx 10^{17}\rm{cm^{-2}}$ as predicted by the $c_s t_{\rm cool}$ ansatz (section \ref{sec:rclsim}). 

    \item[\textbf{3.}] Despite the initial setup of uniform and parallel magnetic fields, the varation of density structures within the initial fractal cloud lead to field lines tangled and draped around the later-formed cloudlets. Variation of magnetic field strength in the simulations shows magnetic fields alter the morphology of the cold gas. It can suppress thermal conduction and mixing with the hot winds, preserving strong temperature gradients. 

    \item[\textbf{4.}] We present a new multiphase model of the CGM based on the implications of small cloudlet sizes. We estimate the number of cloudlets in the inner region of the CGM is in the order of $\sim10^{6-9}$ with an average inter-cloudlet spacing of $\sim30$ cloudlet radii. Although the volume filling factor is small ($10^{-3}$), the covering fraction can reach unity if integrated over only a short $\sim$kpc length scale. This provides a natural explanation for the ubiquitous detection of cold gas in the inner galactic halo. 

    \item[\textbf{5.}]  We also discuss the spatial extent of cold, warm and hot gas as implied by the model. Regardless of the onset of thermal instability, the model naturally predicts that cold cloudlets can only exist up to $\sim$100 kpc as the condensates are compressed to a density regime where it is favored by low ions (e.g., Mg\,II, Si\,II, C\,II). For larger distances, the pressure of the hot gas is relatively lower. Intermediate (e.g., C\,IV) and high ions (O\,VI) will, therefore, still thrive in these conditions (i.e., lower density and higher temperature).

\end{enumerate}

The model as described in this paper represents a multiphase state of the CGM. The inner region is filled with cold, dense, and small gas clouds, resembling terrestrial mist or clouds. This stretches the relevant scales from 0.1 pc to hundreds of kpc in the CGM. The challenge of galaxy formation and evolution remains on the self-regulation of heating and cooling. This regulation and interplay between feedback and cooling of the hot halo gas can determine whether a galaxy will be star-forming or quenched and remains quiescent over a long timescale. The state of the galaxy should also be understood via the steady state of the CGM. Self-consistent galaxy simulations must incorporate the process of cold gas fragmentation with the interplay of star formation and feedback processes. 

\section*{Acknowledgments}

CL is grateful for many stimulating discussions with Andrey Kravtsov and Nick Gnedin. CL thanks Ellen Zweibel and Don York for their comments on the early draft. CL thanks Mike McCourt, Peng Oh, and Joop Schaye for sharing their insights during the ``What matters around galaxies" workshop in the summer of 2017 at Durham, England. CL is grateful for the discussions with Claude-Andr\'e Faucher-Gigu\`ere, Jonathan Stern, and Zachary Hafen during the Chicago-Northwestern joint group meetings.

CL was supported by NASA Headquarters under the NASA Earth and Space Science Fellowship Program - Grant NNX15AR86H. CL is partially supported by a NASA ATP grant NNH12ZDA001N, NSF grant AST-1412107, and by the Kavli Institute for Cosmological Physics at the University of Chicago through grant PHY-1125897 and an endowment from the Kavli Foundation and its founder Fred Kavli. This work has also been partially supported by NSF grants under  DGE-1144082, DGE-1746045 and AST-0709181. The simulations presented in this paper have been carried out using the Midway cluster at the University of Chicago Research Computing Center, which we acknowledge for support.





\bibliographystyle{mnras}
\bibliography{ms} 








\bsp	
\label{lastpage}
\end{document}